\documentclass[aip,jcp,amsmath,amssymb,preprint]{revtex4-1}
\usepackage[utf8]{inputenc} 

\usepackage{geometry} 
\usepackage{graphicx} 
\usepackage{color}

\usepackage{booktabs} 
\usepackage{array} 
\usepackage{paralist} 
\usepackage{verbatim} 
\usepackage{subfig} 
\usepackage{hyperref}

\newcommand*{\citen}[1]{%
  \begingroup
    \romannumeral-`\x 
    \setcitestyle{numbers}%
    \cite{#1}%
  \endgroup   
}

\makeatletter
\newcommand*{\rom}[1]{\expandafter\@slowromancap\romannumeral #1@}
\makeatother

\begin{document}
\title{Electronic friction near metal surfaces: a case where molecule-metal couplings depend on nuclear coordinates}

\author{Wenjie Dou$^{1}$ and Joseph E. Subotnik}
\affiliation{$^{1}$ Department of Chemistry, University of Pennsylvania, Philadelphia, Pennsylvania 19104, USA }

\begin{abstract}
We derive an explicit form for the electronic friction as felt by a molecule near a metal surface for the general case that molecule-metal couplings depend on nuclear coordinates.  Our  work generalizes a  previous study by von Oppen \textit{et al} [Beilstein Journal of Nanotechnology, 3, 144, 2012], where we now go beyond the Condon approximation (i.e. molecule-metal couplings are not held constant).  Using a non-equilibrium Green's function formalism in the adiabatic limit, we show that fluctuating metal-molecule couplings lead to new frictional damping terms and random forces, plus a correction to the potential of mean force.  Numerical tests are performed and compared with a modified classical master equation; our results indicate that violating the Condon approximation can have  a large effect on dynamics.  
\end{abstract}

\maketitle

\section{introduction}
The coupled electron-nuclear dynamics of molecules near metal surfaces underlie many electrochemical phenomena, and have gained a lot of interest recently. For example, vibrational promoted electron transfer and vibrational relaxation for NO molecules scattering from gold surface have been reported \cite{science2000, wodtke} experimentally and followed up by many theoretical studies \cite{shenvi:2009:iesh, shenvi:2009:science}. Coupled electron-nuclear dynamics also play an important role in molecular junctions, and are presumed to account for a great deal of exotic phenomena, including inelastic scattering signatures \cite{inelastic, prldata, bowler2007, mishaScience2008}, hysteresis \cite{Galperinnano, rater2005review, hysertesisJPCC, switching2006}, vibrational heating and cooling \cite{prbnitzan, heating, thossinstabilities, cooling}. 

In the presence of metal surfaces, a manifold of electronic degrees of freedom (DoFs) take part in the dynamics, such that  no simple solution is obvious. One attempt to simplify the dynamics is to treat the electronic bath as a source of friction for the nuclear DoFs. \cite{tully1980, adelman1976} Decades ago, Head-Gordon and Tully (HGT) derived a model for electronic friction based on a smeared view of derivative couplings in the adiabatic limit. \cite{FrictionTully}
Such a formalism has been used successfully in many systems \cite{tullyfrictionexpPRL1996, tullyappPCCP2011, tullyfrictionreview} and yet  apparently fails in other cases. \cite{wodtke, wodtkereview2004}  Following a non-equilibrium Green's function and scattering matrix approach, von Oppen and co-workers have given an alternative formalism  for electronic friction, one which can be generalized to the out of equilibrium case. \cite{beilstein, vonOppenPRB} Similar results are reported from other approaches. \cite{brandbyge, Mozyrsky, lvPRBfriction} In a recent paper, we showed that a classical master equation gives the same friction as von Oppen's model, provided the level broadening can be discarded. \cite{paperIV} In that same paper, we also showed the connection between the HGT and von Oppen's model of friction, both of which share several common features as well as some differences. 

It should be emphasized that von Oppen's friction model relies on a constant molecule-metal coupling. For many systems such as gas molecule scattering from metal surface problem,  molecule-metal  couplings clearly depend on nuclear coordinates. In this paper, we will generalize von Oppen's model to include such non-Condon effects, and give a compact form of electronic friction in general. Interestingly, similar results for friction have previously been derived using purely time-dependent formalisms  (without any nuclear motion) \cite{Mizielinski2005, Mizielinski2007, mishaPRBfriction}; in fact, our final form of friction can be viewed as a generalization of the HGT model to nonzero temperature; see Appendix \ref{app:HGT}. In the present article, we will go beyond previous work by showing that  non-Condon frictional terms come along with additional non-Condon contributions to the random force. At equilibrium, the fluctuation-dissipation theorem is satisfied automatically. Finally and perhaps most importantly, one finds non-Condon effects change the potential of mean force and these changes can be very large.

One shortcoming of our analysis here is that we restrict ourselves to the adiabatic regime, whereby we assume the nuclear motion is much slower than the electronic motion. Now, over the past year, we have argued that it is possible to construct a broadened classical master equation valid in both non-adiabatic and adiabatic regimes. \cite{paperII,paperIII,paperIV} That being said, we will show below that incorporating non-Condon effects is nontrivial in practice and can be done most easily with only a partial treatment (whereby only the contribution to the mean force is incorporated). Numerical tests will show that incorporating such a contribution to the mean force can dramatically affect the dynamics.

We organize this paper as follows. In Sec. \ref{sec:theory}, we introduce our model, and use an adiabatic expansion to derive the correct form of friction. In Sec. \ref{sec:cme}, we introduce our modified classical master equation.  
We discuss the results in Sec. \ref{sec:results} and conclude in Sec. \ref{sec:con}. In the Appendix, we provide additional details for all derivations  as well as show an explicit connection between the HGT model and our analysis.

\section{Theory}  \label{sec:theory}
\subsection{Anderson-Holstein model}
We consider a generalized Anderson-Holstein (AH) model, where an impurity level (with creation [annihilation] operator $d^+$ [$d$]) couples both to a set of nuclear degrees of freedom  and a manifold of electronic states (with creation [annihilation] operator $c_k^+$ [$c_k$]).
\begin{eqnarray} \label{eq:ah1}
 {H} &=& {H}_s+  {H}_b+  {H}_c,  \\
\label{Hs}
 {H}_s &=& h(x)  {d}^+ {d} + \frac{p^2}{2m} + U(x), \\
 {H}_b &=& \sum_k \epsilon_k  {c}_k^+  {c}_k, \\
 \label{eq:ah4}
 {H}_c &=& \sum_k {V_k}(x) (  {d}^+  {c}_k +  {c}_k^+  {d} ). 
\end{eqnarray}
Here, without loss of generality, we have considered only a single nuclear DoF ($x$, $p$); for more general results, see Appendix \ref{app:manyDoF}.

The main difference between our Hamiltonian (Eqs. \ref{eq:ah1}-\ref{eq:ah4}) and the Hamiltonian in Ref. \citen{beilstein} is that, in our model, the molecule-metal coupling $V_k(x)$ depends on nuclear coordinates, which will become the source of new frictional damping forces and random forces. Below, to simplify our discussion, we will assume $V_k (x)$ is independent of \textit{k}, and we will apply the wide band approximation (such that the real part of the retarded self energy $\Sigma^R(\epsilon, x)$ vanishes, and the imaginary part ($-\Gamma(x)/2$) is energy independent), 
\begin{eqnarray} \label{eq:Gamma}
\Sigma^R(\epsilon,  x) \equiv \sum_k \frac{V_k^2 (x)}{\epsilon-\epsilon_k+i\eta}= -i\pi \sum_k V_k^2  (x)\delta(\epsilon-\epsilon_k)= -i\pi V^2 (x) \rho(\epsilon)\equiv -i\Gamma (x)/2.  \nonumber \\
\end{eqnarray}
In the above equation, $\eta$ is a positive infinitesimal.  


In our discussion, we will consider only classical nuclei. If $\omega$ is a frequency for the nuclear motion as estimated by $\omega=\sqrt{\partial_x^2 U/m}$, we assume $kT \gg \hbar\omega$. Then, Newtonian mechanics can be applied for the classical nuclei,  
\begin{eqnarray} \label{eq:eomx}
-m\ddot x &=& \partial_x U  + \partial_x h \: d^+ d + \sum_k \partial_x   {V_k}  (  {d}^+  {c}_k +  {c}_k^+  {d} ) \nonumber \\
&=&\partial_x U + \partial_x h  \: d^+ d + \frac{\partial_x \Gamma } {2\Gamma (x)} \sum_k {V_k}  (x)  (  {d}^+  {c}_k +  {c}_k^+  {d} ). 
\end{eqnarray}
The last equality in the above equation comes from the assumptions that $V_k(x)$ is independent of $k$, such that $ \frac{\partial_x {V_k} } {{V_k}(x)}  = \frac{\partial_x \Gamma} {2\Gamma  (x)}$ (see Eq. \ref{eq:Gamma}). 

In Eq. \ref{eq:eomx}, the nuclear motion is highly coupled with the electronic DoFs. For a useful frictional model, we would like to transform Eq. \ref{eq:eomx}  into a closed set of Langevin equations for purely nuclear DoFs, 
\begin{eqnarray} \label{eq:EF-LD}
-m\ddot x =\partial_x U - F (x) +\gamma (x) \dot{x} + \delta f( x,t), 
\end{eqnarray}
where $F(x)$, $\gamma (x)$ and $\delta f(x, t)$ are the mean force, frictional damping coefficient  and random force that the nuclei experience as caused by the electronic DoFs.  In the adiabatic limit, where the electronic motion is much faster than the nuclear motion, $\Gamma \gg \hbar\omega$, such a transformation is possible. We will show below that is natural to write: 
\begin{eqnarray} \label{eq:decompose}
F(x) &=& F_1(x) + F_2 (x),  \label{eq:F}\\
\gamma(x) &=& \gamma_1(x) + \gamma_2(x) + \gamma_3 (x)+ \gamma_4(x),  \label{eq:gamma} \\
D(x) &=& D_{11}(x) + D_{12}(x) + D_{21}(x) + D_{22}(x),   \label{eq:corrDtotalso} 
\end{eqnarray}
where $D(x)$ is the correlation function of the random force 
\begin{eqnarray}
\langle \delta f(x, t) \delta f(x,t') \rangle &=& D(x) \delta(t-t'). 
\end{eqnarray}
All terms above will be defined below. 

\subsection{Green's functions}
We will now show how to transform Eq. \ref{eq:eomx} into Eq. \ref{eq:EF-LD} using the language of Green's functions. To do so, we require a few preliminary definitions. 
\subsubsection{Equilibrium (Frozen) Green's functions}
Without nuclear motion, the Hamiltonian in Eqs. \ref{eq:ah1}-\ref{eq:ah4}  is the trivial resonant level model and can be solved with equilibrium Green's functions\cite{mahan} that assume fixed nuclei and depend only on the time difference:
\begin{eqnarray} \label{eq:Gretard}
G^R(t-t',x) &\equiv&  -\frac{i}{\hbar} \theta (t-t') \langle \{ d(t), d^+(t') \} \rangle_{x},  \\
G^<(t-t',x) &\equiv& \frac{i}{\hbar} \langle d^+(t') d(t) \rangle_{x}. \label{eq:Gless}
\end{eqnarray}
Here $\{ , \}$ denotes the anti-commutator. 
Frozen, equilibrium Green's functions are most naturally expressed in the energy domain, $G(t-t',x)=\int \frac{d\epsilon}{2\pi\hbar} G(\epsilon,x) e^{-i\epsilon (t-t') /\hbar} $  as follows:
\begin{eqnarray} \label{eq:GretardEx}
G^R(\epsilon,x) &=& \frac{1}{\epsilon - h(x) -\Sigma^R}, \\
G^<(\epsilon,x) &=& i A(\epsilon, x) f(\epsilon),  \label{eq:GretardEx2}  
\end{eqnarray}
where $A(\epsilon, x)$ is the spectral function, 
\begin{eqnarray}
 A(\epsilon,x) = \frac {\Gamma(x)} {(\epsilon -h(x))^2 + (\Gamma(x)/2)^2}, 
\end{eqnarray} 
 and $f(\epsilon) \equiv \frac1{\exp(\beta(\epsilon-\mu) )+1}$ is the Fermi function. 

\subsubsection{Nonequilibrium Green's functions}
Now, when nuclear motion is included, frozen Green's functions can be invoked only if nuclear motion is infinitesimally slow, such
that the electrons have no memory of any nuclear motion and $x(0)$ is sampled from a static distribution.
More generally, we can define time-dependent nonequilibrium Green's functions as follows:
\begin{eqnarray} \label{eq:fulGretard}
\tilde G^R(t, t') & \equiv& -\frac{i}{\hbar} \theta (t-t') \langle \{ d(t), d^+(t') \} \rangle_{x(t)}\\
\tilde G^< (t,t') & \equiv & \frac{i}{\hbar} \langle d^+(t') d(t) \rangle_{x(t)}, \label{eq:fulGless}
\end{eqnarray}
Here, $\langle ...\rangle_{x(t)}$ implies average over electronic DoFs for a given trajectory $x(t)$.
Whereas $G(t-t')$ does not depend on the velocity of the nuclei at time $t$,  $\tilde{G}(t,t')$ does depend on such velocity.  (Formally, we should write $\tilde{G} (t, t',[x(t)])$, but this notation would be very cumbersome.) 

 Note that $G (t-t')$ and $\tilde{G} (t,t')$ are only one element of a bigger set of Green's functions. Below we will also need 
\begin{eqnarray}
\tilde G_{d,k}^< (t,t') &\equiv& \frac{i}{\hbar} \langle c_k^+(t') d(t) \rangle_{x(t)}. \label{eq:fulGkless}
\end{eqnarray}

Using these definitions, we can separate the operator on the right hand side of Eq. \ref{eq:eomx} into an average part and a random part. For example, for the $\partial_x h (d^+d)$ term, we write $d^+d = \langle d^+d \rangle + (d^+d -  \langle d^+d \rangle)$. 
Eq. \ref{eq:eomx} then becomes 
\begin{eqnarray}\label{eq:starter}
-m\ddot x =\partial_x U + \partial_x h (- i \hbar \tilde{G}^<(t,t)) +  \frac{\partial_x \Gamma} {2\Gamma} \sum_k {V_k} 2\Re(-i\hbar\tilde{G}^<_{d, k} (t,t) ) + \delta f(x, t), 
\end{eqnarray}
where $\delta f(x, t)$ is the random force, 
\begin{eqnarray}
\delta f(x, t) &=& \delta f_1(x, t) + \delta f_2(x, t), \\
\delta f_1(x, t) &=&  \partial_x h (d^+ d + i\hbar \tilde{G}^<(t,t) ), \label{eq:deltaf1} \\
\delta f_2(x, t) &=& \frac{\partial_x \Gamma}{2\Gamma} \sum_k    {V_k} \big(  {d}^+  {c}_k +  {c}_k^+  {d} + 2\Re(i\hbar\tilde{G}^<_{d, k}(t,t)) \big). \label{eq:deltaf2}
\end{eqnarray}

Below we will calculate explicit forms for all terms in Eq. \ref{eq:starter} in the limit of slow nuclear motion using a gradient expansion of the Green's functions. 
Because non-equilibrium Green's functions are nonstandard in chemistry, we will refer the reader to Ref. \citen{JH1877notes} for the relevant background when necessary.

\subsubsection{Wigner transformation} 
Below, to perform a gradient expansion, we will require frequent use of a Wigner transformation which
allows us to separate fast electronic motion from slow nuclear motion. The Wigner transformation of $C(t_1,t_2)$ is defined as
\begin{eqnarray}
C^W(t,\epsilon)=\int d\tau e^{i\epsilon \tau/\hbar} C(t+\tau/2, t-\tau/2). 
\end{eqnarray}
As is well known \cite{Tannor}, the Wigner transformation of a convolution $C(t_1, t_2) = \int dt_3 \: A(t_1, t_3) B(t_3, t_2)$
can be expressed with a ``Moyel operator'' as:
\begin{eqnarray}\label{moyel}
C^W (t, \epsilon) = \exp\left[\frac{i\hbar}{2}\left(\partial_{\epsilon}^A \partial_t^B - \partial_{\epsilon}^B \partial_t^A \right) \right] A^W B^W \approx A^W B^W  + \frac{i\hbar}{2}  \partial_{\epsilon} A^W \partial_t B^W - \frac{i\hbar}{2}  \partial_{\epsilon} B^W \partial_t A^W.  \nonumber \\
\end{eqnarray}
On the far right hand side of Eq. \ref{moyel}, the expansion is  correct to order $\hbar$. Eq. \ref{moyel} is sometimes called a gradient expansion.

\subsubsection{Notation}
 From now on, unless otherwise noted, we will use $\tilde{G}$ ($G$) to denote $\tilde{G}(t, \epsilon, [x(t)])$ ($G(\epsilon, x)$). In other words, for frozen Green's functions, we will work almost always in the energy domain (rather than the time domain). For non-equilibrium Green's functions, we will work almost exclusively with the Wigner transformation. When we want to work in the time domain explicitly, we will write $\tilde{G}(t,t')$ ($G(t-t')$).

\subsection{Gradient expansion}
\subsubsection{Gradient expansion of $\tilde{G}^R(t,t')$}
We begin by analyzing
the retarded Green's function $\tilde{G}^R(t,t')$.
In Ref. \citen{beilstein}, von Oppen \textit{et al} showed that, for the case of a single impurity level and constant $\Gamma$, the full $\tilde{G}^R$ is equal to the frozen $G^R$ up to the linear order in the velocity of the nuclei, $\tilde{G}^R=G^R$. 
Let us now show that, $\tilde{G}^R=G^R$ still holds when $\Gamma$ depends on nuclear coordinates. 

To demonstrate the equivalence, following von Oppen \textit{et al}, note that 
the equation of motion for the retarded Green's function (as a function of $t'$) is given by
\begin{eqnarray} \label{eq:eomGR}
-i\hbar \partial_{t'} \tilde{G}^R(t, t')= \delta(t-t') + \int dt_1 \tilde{G}^R(t, t_1) \Sigma^R(t_1, t') +  \tilde{G}^R(t, t') h. 
\end{eqnarray}
We emphasize that the derivative of the fully time-dependent Green's function   $\tilde{G}^R(t, t')$ 
(Eq. \ref{eq:fulGretard}) with respect to $t'$ is the same as the derivative with respect to $t'$ of the frozen
Green's function $G^R(t-t')$ (Eq. \ref{eq:Gretard}). This statement is not true for the derivative with respect to $t$.

After a Wigner transformation (and a gradient expansion), Eq. \ref{eq:eomGR} becomes
\begin{eqnarray} \label{eq:WGR}
\tilde{G}^R ( \epsilon - \Sigma^R -h ) = 1 + \frac{i\hbar}2 \partial_{\epsilon}  \tilde{G}^R \partial_{t} h + \frac{i\hbar}2 \partial_{\epsilon}  \tilde{G}^R \partial_{t} \Sigma^R + \frac{i\hbar}2 \partial_{t}  \tilde{G}^R (1-\partial_{\epsilon} \Sigma^R). 
\end{eqnarray}
and dividing by $(G^R)^{-1} = \epsilon - \Sigma^R -h$, we find:
\begin{eqnarray} \label{eq:WGR2}
\tilde{G}^R = G^R + \frac{i\hbar}2 \left[ \partial_{\epsilon}  \tilde{G}^R \partial_{t} h + \partial_{\epsilon}  \tilde{G}^R \partial_{t} \Sigma^R +  \partial_{t}  \tilde{G}^R (1-\partial_{\epsilon} \Sigma^R) \right]G^R. 
\end{eqnarray}
At this point, the only difference between our treatment of the problem and von Oppen's derivation
in Ref. \citen{beilstein} is that, in our case, since $\Sigma^R$ depends on $x$, $\partial_t \Sigma^R\neq0$. Instead, note
that $\partial_t=\dot{x} \partial_x$, so that all of the terms in brackets
on the right hand side of Eq. \ref{eq:WGR2}  are already first order in velocity.  Thus, inside the brackets,
to first order in velocity we can 
approximate $\tilde{G}^R = G^R$. Thus, we find:
\begin{eqnarray}
\tilde{G}^R &\approx & G^R + \frac{i\hbar}2 \left[ \partial_{\epsilon}  {G}^R \partial_{t} h + \partial_{\epsilon}  {G}^R \partial_{t} \Sigma^R +  \partial_{t}  {G}^R (1-\partial_{\epsilon} \Sigma^R) \right]G^R\nonumber \\
 &=& G^R + \frac{i\hbar}2 [\partial_{\epsilon}  {G}^R \partial_{t} h +  \partial_{\epsilon} {G}^R \partial_{t} \Sigma^R+ ( \partial_{t} h +  \partial_{t} \Sigma ^R) ( {G}^R)^2 (1-\partial_{\epsilon} \Sigma^R) ] G^R \label{app1}\\
&=&  G^R + \frac{i\hbar}2 [\partial_{\epsilon}  {G}^R \partial_{t} h +  \partial_{\epsilon}  {G}^R \partial_{t} \Sigma^R- ( \partial_{t} h +  \partial_{t} \Sigma ^R) \partial_{\epsilon}  G^R ] G^R = G^R. 
\end{eqnarray}
Here, we have differentiated $G^R=1/(\epsilon- h- \Sigma^R)$ (Eq. \ref{eq:GretardEx}), and used  the fact that $\partial_{t} G^R = ( \partial_{t} h +  \partial_{t} \Sigma ^R) (G^R)^2$, and $ \partial_{\epsilon} G^R = - (1-\partial_{\epsilon} \Sigma^R) (G^R)^2$.  This proves our hypothesis that
$\tilde{G}^R=G^R$ to first order in $\dot{x}$.

\subsubsection{Gradient expansion of $\tilde{G}^< $}
We are now ready to perform a gradient expansion of the lesser Green's function $\tilde{G}^<$ (as it appears in Eq. \ref{eq:starter}).
We begin by considering
the Langreth relation $\tilde{G}^< (t, t')  = \int dt_1 dt_2 \tilde{G}^R (t, t_1) \Sigma^< (t_1, t_2) \tilde{G}^A(t_2, t') $ 
(Eq. 39  of Ref. \citen{JH1877notes}) for the Dyson equation of the contour-ordered Green's function.  
 We perform a Wigner transformation using Eq. \ref{moyel} two times, and we find:
\begin{eqnarray} \label{eq:adiaGless}
\tilde{G}^< \approx G^< &+& \frac{i\hbar}2 \partial_t h [ \partial_{\epsilon} G^< G^A -  G^< \partial_{\epsilon} G^A -  G^R \partial_{\epsilon} G^< + \partial_{\epsilon}  G^R G^< ]  \nonumber \\
&-& \frac{i\hbar}2 \partial_t \Sigma^R [ \partial_{\epsilon} G^< G^A - G^<  \partial_{\epsilon} G^A + G^R \partial_{\epsilon} G^<  - \partial_{\epsilon} G^R  G^<]   \nonumber \\
&+& \frac{i\hbar}2 \partial_t \Sigma^< [ \partial_{\epsilon} G^R G^A - G^R \partial_{\epsilon} G^A ] . 
\end{eqnarray}
Here, we have used the same Langreth relation  for the frozen lesser Green's function, $G^<= G^R\Sigma^< G^A$, and 
we have  differentiated $G^A=( G^R )^* = 1/(\epsilon- h + \Sigma^R)$ , such that $\partial_t G^A= ( \partial_{t} h -  \partial_{t} \Sigma ^R) (G^A)^2$. Note that we have replaced $\tilde{G}^{R/A}$ with ${G}^{R/A}$ on the right hand side of Eq. \ref{eq:adiaGless}, which is correct to the first order in velocity.

When we examine Eq. \ref{eq:adiaGless}, the frozen retarded Green's function $G^<$ gives a mean force $F_1(x)$ on the nuclei  as seen in Eq. \ref{eq:F} (and using Eq. \ref{eq:GretardEx2}), 
\begin{eqnarray} \label{eq:F1}
F_1 (x)= -\partial_x h \int \frac{d\epsilon}{2\pi} (-iG^<) = -\partial_x h \int \frac{d\epsilon}{2\pi}  A (\epsilon, x) f(\epsilon) . 
\end{eqnarray} 
Knowing $\partial _t =\dot{x} \partial_x $,  the second set of terms on the right hand side of Eq. \ref{eq:adiaGless} gives a friction term $\gamma_1(x)$ (Eq. \ref{eq:gamma}), 
\begin{eqnarray} \label{eq:gamma1raw}
\gamma_1 (x) = \hbar(\partial_x h)^2 \int \frac{d\epsilon}{2\pi} [ \partial_{\epsilon} G^< (G^A-G^R)]  . 
\end{eqnarray}
In the above equation, we have used integration by parts, $\int {d\epsilon} X \partial_{\epsilon} Y  = -\int d\epsilon  Y \partial_{\epsilon} X$. Below we will require this trick repeatedly.  

The last two terms in Eq. \ref{eq:adiaGless} give another friction term $\gamma_2(x)$, 
\begin{eqnarray} \label{eq:gamma2raw}
\gamma_2  (x) = (\partial_x h \partial_x \Gamma)   \frac{i\hbar}2 \int \frac{d\epsilon}{2\pi} [\partial_{\epsilon} G^< (G^A+G^R)+ (\partial_{\epsilon} G^R G^A-  G^R \partial_{\epsilon} G^A) f(\epsilon) ] , 
\end{eqnarray}
where we have used $\Sigma^< = i \Gamma f(\epsilon)$. 
$\gamma_1(x)$ and $\gamma_2(x)$ can be recast into a compact form with all frozen Green's functions known explicitly (see Appendix \ref{app:a}), 
\begin{eqnarray}
\gamma_1 (x) &=&   - (\partial_x h)^2 \frac{\hbar}2 \int \frac{d\epsilon}{2\pi} A^2(\epsilon,x) \partial_{\epsilon} f(\epsilon) \nonumber \\
&=&  - (\partial_x h)^2 \frac{\hbar}2 \int \frac{d\epsilon}{2\pi} \left( \frac {\Gamma(x)} {(\epsilon -h(x))^2 + (\Gamma(x)/2)^2} \right)^2\partial_{\epsilon} f(\epsilon),  \label{eq:gamma1}  \\
\gamma_2 (x) &=& - \frac{\hbar} 2 (\partial_x h \partial_x \Gamma)   \int \frac{d\epsilon}{2\pi}  \frac{(\epsilon-h(x)) A^2(\epsilon,x) }{\Gamma(x)} \partial_{\epsilon} f(\epsilon). \label{eq:gamma2}
\end{eqnarray}

\subsubsection{ Gradient expansion of $\tilde{G}^<_{d, k}$ }
Finally, we evaluate the last Green's function $\tilde{G}^<_{d, k}$ appearing in Eq. \ref{eq:starter}.
Again, we use the Langreth trick (Eq. 54 of Ref. \citen{JH1877notes})
for the Dyson equation. We find:
\begin{eqnarray} \label{eq:GradientGdk}
\tilde{G}^<_{d, k} (t, t') =  \int dt_1 \tilde{G}^R(t, t_1) {V}_k g_k^< (t_1, t') +  \tilde{G}^< (t, t_1) {V}_k g_k^a(t_1, t'). 
\end{eqnarray}
Here, $g_k(t,t') = G_{kk}^0(t,t')$ is the noninteracting Green's function for an electron in the lead, and is easily written in the energy domain, $g_k (t, t') =  \int \frac{d\epsilon}{2\pi\hbar} \: g_k(\epsilon ) e^{-i \epsilon (t-t')/\hbar } $, with
\begin{eqnarray}
g_k^a (\epsilon) &= & \frac{1} {\epsilon-\epsilon_k - i \eta}  \\
g_k^< (\epsilon) &=& i 2\pi \delta (\epsilon-\epsilon_k) f(\epsilon)  
\end{eqnarray}

As above, we perform a Wigner transformation, and using the fact that $\partial_t g^< (\epsilon) = \partial_t g_k^a (\epsilon) = \partial_{\epsilon} V_k = 0$, we find that, to the first order in velocity:
\begin{eqnarray} \label{eq:adiaGdk}
\tilde{G}^<_{d, k} 
&\approx& (G^R V_k g^<_k + \tilde{G}^<V_k g^a_k )  -  \frac{i\hbar}2  {V}_k  (   \partial_t G^R \partial_{\epsilon} g_k^< +  \partial_t G^< \partial_{\epsilon} g_k^a) \nonumber \\
&+&  \frac{i\hbar}2 (   \partial_{\epsilon} {G}^R  g_k^< -  {G}^R  \partial_{\epsilon} g_k^<  +    \partial_{\epsilon}  {G}^<  g_k^a -  {G}^<  \partial_{\epsilon} g_k^a )   \partial_t  {V}_k  . 
\end{eqnarray} 

Let us now discuss the individual terms on the right hand side of Eq. \ref{eq:adiaGdk}. 
The frozen $G^R V_k g^<_k$ term gives a second contribution to the mean force $F_2(x)$ (in Eq. \ref{eq:decompose}), 
\begin{eqnarray}\label{eq:blowup}
F_2(x)= -\frac{\partial_x \Gamma} {2\Gamma(x)} \sum_k V_k  \int \frac{d\epsilon}{2\pi} 2\Re(-i G^R V_k g^<_k ) . 
\end{eqnarray}

Using Eq. \ref{eq:corrGlessgreat} in the Appendix, one can write down an explicit form for $F_2(x)$.
As discussed in detail in the Appendix of Ref. \citen{paperVI}, the integral in Eq. \ref{eq:blowup} will blow up if we integrate from $-\infty$ to $\infty$. Thus, as in Ref. \citen{paperVI}, we introduce a band width ($-W$, $W$) to evaluate $F_2(x)$ (while still insisting that $W \gg \Gamma$ 
so that we can ignore dynamical effects beyond the wide-band limit). The final answer is:
\begin{eqnarray} \label{eq:F2}
F_2(x) =-\frac{\partial_x \Gamma} {\Gamma(x)}  \int_{-W}^W \frac{d\epsilon}{2\pi}   (\epsilon-h(x)) A(\epsilon, x)  f(\epsilon) . 
\end{eqnarray}
The contribution of the term $\tilde{G}^<V_k g^a_k$  (in Eq. \ref{eq:adiaGdk}) to the force (Eq. \ref{eq:F}) is zero because $\Re \sum_k V_k^2 g^a_k=0$ (i.e. the wide band limit). The second and third set of terms on the right hand side of Eq. \ref{eq:adiaGdk} make further contributions to the frictional damping 
($\gamma_3(x)$ and $\gamma_4(x)$ in Eq. \ref{eq:gamma}).  
See Appendix \ref{app:a} for details. We find:
\begin{eqnarray} \label{eq:gamma3}
\gamma_3 (x) &=&  -\frac{\hbar \partial_x \Gamma} {2\Gamma(x)} \sum_k  {V}_k^2(x)   \int \frac{d\epsilon}{2\pi}  \Re (   \partial_x G^R \partial_{\epsilon} g_k^< +  \partial_x G^< \partial_{\epsilon} g_k^a) \nonumber \\
&=& -  \frac{\hbar (\partial_x \Gamma)^2} 4 \int \frac{d\epsilon}{2\pi} (\frac{A(\epsilon, x)}{\Gamma(x)}- \frac{A^2(\epsilon, x)}2) \partial_{\epsilon} f(\epsilon) \nonumber \\
&&-  \frac{ \hbar \partial_x \Gamma \partial_x h  } 2 \int \frac{d\epsilon}{2\pi} \frac{A^2(\epsilon, x)}{\Gamma(x)} (\epsilon-h(x)) \partial_{\epsilon} f(\epsilon) ,  \\
\label{eq:gamma4}
\gamma_4 (x) &=& -\frac{\hbar\partial_x \Gamma} {2\Gamma(x)} \sum_k   {V}_k(x)  \partial_x   {V}_k  \int \frac{d\epsilon}{2\pi} 2\Re (   G^R \partial_{\epsilon} g_k^< +  G^< \partial_{\epsilon} g_k^a) \nonumber  \\
&=&  -  \frac{\hbar(\partial_x \Gamma)^2} {4}   \int \frac{d\epsilon}{2\pi}  \frac{A(\epsilon, x)}{\Gamma(x)} \partial_{\epsilon} f(\epsilon) . 
\end{eqnarray}

\subsection{Fluctuation-dissipation theorem}
Now we will evaluate the correlation functions of the random force $\delta f(x, t)=\delta f_1(x, t)+\delta f_2(x, t)$ (Eqs. \ref{eq:deltaf1}-\ref{eq:deltaf2}). In the adiabatic limit, we would like the correlation function of the random force to be Markovian, 
\begin{eqnarray}
\langle \delta f_i(x, t) \delta f_j (x, t') \rangle = D_{ij}(x) \delta (t-t'), \: i,j=1,2. 
\end{eqnarray}
We start by applying Wick's theorem:      
\begin{eqnarray}
\langle \delta f_1(x, t) \delta f_1(x, t')\rangle  &=&  \hbar^2 (\partial_x h)^2 \tilde{G}^> (t,t') \tilde{G}^< (t',t), \label{eq:corr1}\\
\langle \delta f_1(x, t) \delta f_2 (x, t')\rangle &=& \hbar^2  \frac {\partial_x h \partial_x \Gamma}  {2\Gamma}  2 \Re \sum_k   {V}_k  \tilde{G}_{d,k}^> (t,t') \tilde{G}^< (t',t) ,\\
\langle \delta f_2(x, t) \delta f_1 (x, t')\rangle &=& \hbar^2  \frac {\partial_x h \partial_x \Gamma}  {2\Gamma}  2\Re  \sum_k   {V}_k \tilde{G}^> (t,t') \tilde{G}_{d,k}^< (t',t), \\
\langle \delta f_2(x, t) \delta f_2(x, t')\rangle &=& \hbar^2 \left( \frac{\partial_x \Gamma } {2\Gamma} \right ) ^2 2   \Re\sum_{k,k'}   {V}_k   {V}_{k'} \tilde{G}_{k,d}^> (t,t')  
\tilde{G}_{k',d}^< (t',t)      \nonumber     \\
&+&\hbar^2 \left( \frac{\partial_x \Gamma } {2\Gamma} \right ) ^2 2   \Re\sum_{k,k'}   {V}_k   {V}_{k'}   \tilde{G}_{k,k'}^> (t,t') \tilde{G}^< (t',t) .  \label{eq:corr4}
\end{eqnarray}
In the above equations, $\tilde{G}^>$ is the greater Greens function defined as 
\begin{eqnarray} 
\tilde G^> (t,t') &=&- \frac{i}{\hbar} \langle d(t) d^+(t') \rangle_{x(t)}, \label{eq:fulGgreat}\\
\tilde G_{d,k}^> (t,t') &=&- \frac{i}{\hbar} \langle d (t) c_k^+(t') \rangle_{x(t)}, \label{eq:fulGkgreat} \\
\tilde G_{k,k'}^> (t,t') &=&- \frac{i}{\hbar} \langle c_{k}(t) c_{k'}^+(t') \rangle_{x(t)}. \label{eq:fulGkkgreat} 
\end{eqnarray}

For Markovian dynamics, we must replace the corresponding full Green's functions in Eqs. \ref{eq:corr1}-\ref{eq:corr4} by the frozen Green's functions, so that all Green's functions depend only on $\tau = t - t'$.
In such case, the correlation function can be evaluated explicitly. For instance,  
\begin{eqnarray}
D_{11}  (x) &=& \hbar^2 (\partial_x h)^2 \int d\tau G^>(\tau) G^<(-\tau) \\
&=&\hbar (\partial_x h)^2 \int \frac{d\epsilon}{2\pi} G^>(\epsilon) G^<(\epsilon)  \\
&=& \hbar (\partial_x h)^2 \int \frac{d\epsilon}{2\pi} A^2(\epsilon, x) f(\epsilon)(1-f(\epsilon)). 
\end{eqnarray}

In Appendix \ref{app:b}, we evaluate the other terms. The end results are:
\begin{eqnarray} \label{eq:D1221s}
D_{12} (x) &=&D_{21} (x) =  \hbar \partial_x h   {\partial_x \Gamma}   \int \frac{d\epsilon}{2\pi} \frac{ (\epsilon-h(x)) A^2(\epsilon, x)  }{\Gamma(x)}    f(\epsilon)(1-f(\epsilon)) , \\
\label{eq:D22total}
D_{22}  (x)
&=& \hbar (\partial_x \Gamma)^2 \int \frac{d\epsilon}{2\pi} \frac{(\epsilon-h(x))^2 A^2(\epsilon, x) }{\Gamma^2(x)} f(\epsilon)(1-f(\epsilon)) .
\end{eqnarray}

\subsection{Putting It All Together}
Now we collect together all of the correlation functions for the random force
\begin{eqnarray} \label{eq:corrtotal}
D(x)=\sum_{i,j=1,2} D_{ij}(x) &=& \hbar \int \frac{d\epsilon}{2\pi} \left( \partial_x h + (\epsilon-h(x))\frac{\partial_x \Gamma}{\Gamma(x)} \right)^2 A^2(\epsilon, x)  f(\epsilon)(1-f(\epsilon)) , 
\end{eqnarray}
and friction
\begin{eqnarray} \label{eq:fritotal}
\gamma (x)= \sum_{i=1,4} \gamma_i (x) &=& - \frac{\hbar}{2} \int \frac{d\epsilon}{2\pi} \left( \partial_x h + (\epsilon-h(x))\frac{\partial_x \Gamma}{\Gamma(x)} \right)^2  A^2(\epsilon, x) \partial_{\epsilon} f(\epsilon),   
\end{eqnarray}
We may also evaluate the mean force:
\begin{eqnarray} \label{eq:mftotal}
F (x)=F_1(x)+ F_2(x) = -  \int_{-W}^W \frac{d\epsilon}{2\pi} \left(\partial_x h + (\epsilon-h(x))\frac{\partial_x \Gamma}{\Gamma(x)} \right) A(\epsilon, x)f(\epsilon) . 
\end{eqnarray}
Because $\partial_{\epsilon} f(\epsilon) = -f(\epsilon)(1-f(\epsilon))/kT$, 
we find that our analysis satisfies the fluctuation-dissipation theorem $D(x)=2kT\gamma(x)$. Note that Eq. \ref{eq:fritotal} was reported previously in Refs. \citen{Mizielinski2005, Mizielinski2007, mishaPRBfriction}.


\section{Broadened Classical Master Equation (BCME) and Electron-Friction Langevin Dynamics (EF-LD)}  \label{sec:cme}
In 2015, we analyzed a simple classical master equation (CME) for modeling dynamics in the limit of $\Gamma< kT$ \cite{paperII}  (i.e. assuming weak system-bath coupling), 
and we showed that this CME should be valid both in the non-adiabatic ($\Gamma < \hbar\omega$) and adiabatic ($\Gamma>\hbar\omega$) limit. \cite{paperIII, paperIV} In a more recent paper, we proposed a straightforward, extrapolated approach to incorporate level broadening, 
such that one could extend the range of validity for the CME to include  $\Gamma>kT$. \cite{paperV}  All of our previous work assumed
the Condon approximation, such that $\Gamma(x) = \Gamma$ does not depend on nuclear coordinate $x$.
In this section, we would like to incorporate the extra effect of breaking the Condon approximation
($\partial_x \Gamma$) into our classical master equation (CME). We will show that this can be done, at least partially, by ansatz.

To achieve such a general, broadened classical master equation, we will use the following set of equations
(which constitute a {\em broadened} classical master equation (bCME)), which is valid when $\Gamma$ is a constant:
\begin{eqnarray}  \label{eq:cme1}
\frac { \partial P_0(x,p,t)} {\partial t} &=&  - \frac{p}{m}  \frac {\partial P_0(x,p,t)} {\partial x} + \Big(\partial_x U -f(h(x)) \partial_x h -F_1(x) \Big) \frac {\partial P_0(x,p,t)} { \partial p} \nonumber \\
&-&\frac{\Gamma}{\hbar} f(h(x)) P_0(x,p,t) 
+ \frac{\Gamma}{\hbar} \big(1-f(h(x))\big) P_1(x,p,t), 
\end{eqnarray}
\begin{eqnarray}  \label{eq:cme2}
\frac { \partial P_1(x,p,t)} {\partial t} &=&  - \frac{p}{m}  \frac {\partial P_1 (x,p,t)} {\partial x} +  \Big(\partial_x U + \big(1-f(h(x)) \big) \partial_x h-F_1(x) \Big) \frac {\partial P_1 (x,p,t)} { \partial p}  \nonumber \\
&+& \frac{\Gamma}{\hbar} f(h(x)) P_0(x,p,t) 
- \frac{\Gamma}{\hbar} \big(1-f(h(x))\big) P_1(x,p,t) , 
\end{eqnarray}
where $F_1(x)$ is defined in  Eq. \ref{eq:F1}. 
Eqs. \ref{eq:cme1}-\ref{eq:cme2} are slightly different from our previous work in Ref. \citen{paperV} but basically very similar. 
See Appendix \ref{app:c} for more details. $P_0(x,p,t)$ \big($P_1(x,p,t)$\big) in the above equations is the  probability density for the level in the molecule to be unoccupied (occupied) with nuclei at position $x$ with momentum $p$. We emphasize that Eqs. \ref{eq:cme1}-\ref{eq:cme2} correctly extrapolate between the limits of strong and weak molecule-metal coupling, while always assuming nuclear motion is classical ($kT>\hbar\omega$). To gain intuition for Eqs. \ref{eq:cme1}-\ref{eq:cme2}, 
the most important points to keep in mind are: $(i)$ For small $\Gamma$, $F_1(x)\rightarrow -f(h(x))\partial_x h$, so that Eqs. \ref{eq:cme1}-\ref{eq:cme2} recover the unbroadened CME; \cite{paperIV, paperV} $(ii)$
In the adiabatic limit, following Ref. \citen{paperV}, Eqs. \ref{eq:cme1}-\ref{eq:cme2} yield the same Langevin equation as found by von Oppen {\em et al}\cite{beilstein},
whereby the system evolves adiabatically on a broadened potential of mean force $U_{pmf}$:
\begin{eqnarray} \label{eq:Upmf1}
U_{pmf}(x) = U(x) - \int_{x_0}^x  dx' F_1(x')
\end{eqnarray}
See Ref. \citen{paperV} for instructions on taking the adiabatic limit.

Eqs. \ref{eq:cme1}-\ref{eq:cme2} are very suggestive, as now one can easily incorporate the extra mean force $F_2(x)$ (Eq. \ref{eq:F2}) coming from $\partial_x \Gamma$, 
\begin{eqnarray}  \label{eq:cme3}
\frac { \partial P_0(x,p,t)} {\partial t} &=&  - \frac{p}{m}  \frac {\partial P_0(x,p,t)} {\partial x} +   \Big(\partial_x U - f(h(x)) \partial_x h -F_1(x) -F_2(x) \Big)  \frac {\partial P_0(x,p,t)} { \partial p}  \nonumber \\
&-&\frac{ \Gamma(x)}{\hbar} f(h(x)) P_0(x,p,t) 
+ \frac{\Gamma(x)}{\hbar} \big(1-f(h(x))\big) P_1(x,p,t), 
\end{eqnarray}
\begin{eqnarray}  \label{eq:cme4}
\frac { \partial P_1(x,p,t)} {\partial t} &=&  - \frac{p}{m}  \frac {\partial P_1 (x,p,t)} {\partial x} +  \Big(\partial_x U + \big(1-f(h(x)) \big) \partial_x h -F_1(x) -F_2(x) \Big)  \frac {\partial P_1 (x,p,t)} { \partial p}  \nonumber \\
&+& \frac{\Gamma(x)}{\hbar} f(h(x)) P_0(x,p,t) 
- \frac{\Gamma(x)}{\hbar} \big(1-f(h(x))\big) P_1(x,p,t) . 
\end{eqnarray}
Thus, it is very simple to incorporate any violation of the Condon approximation into a classical master equation, at least regarding the potential of mean force. The new potential of mean force is simply:
\begin{eqnarray} \label{eq:Upmf2}
U_{pmf}(x) = U(x) - \int_{x_0}^x  dx' F_1(x') - \int_{x_0}^x  dx' F_2(x')
\end{eqnarray}

Lastly, to incorporate broadening, we always \cite{paperV} broaden the probability densities $P_0(x,p,t)$ and $P_1(x,p,t)$ as follows,
\begin{eqnarray}
\tilde{P}_0(x,p,t) &=& \big(1- n(h(x))+f(h(x)) \big) P_0(x,p,t) - \big(n(h(x))- f(h(x)) \big)  P_1(x,p,t)   \nonumber  \\ &+& \big( n(h(x))  
- f(h(x)) \big) \big( P_0(x,p,t)+P_1(x,p,t) \big)  \exp \big(-\int_0^t \; dt \; \Gamma(x(t)) \big) \nonumber \\ \\
\tilde{P}_1(x,p,t) &=& \big(1+ n(h(x)) - f(h(x)) \big) P_1 (x,p,t) + \big(n(h(x))- f(h(x)) \big)  P_0 (x,p,t)    \nonumber \\ 
&-& \big(n(h(x)) - f(h(x)) \big) \big( P_0(x,p,t)+P_1(x,p,t) \big)  \exp \big( -\int_0^t\;  dt \; \Gamma(x(t))\big) \nonumber \\
\end{eqnarray}
Here $\tilde{P}_0(x,p,t)$ and $\tilde{P}_1(x,p,t)$ are probability densities that include ad hoc broadening. In the above equations, $n(h(x))$ is the local population defined as 
\begin{eqnarray} \label{eq:nh}
n(h(x)) \equiv \int \frac{d\epsilon}{2\pi} A(\epsilon, x) f(\epsilon) =\int \frac{d\epsilon}{2\pi}  \frac{\Gamma(x)}{(\epsilon-h(x))^2+(\Gamma(x)/2)^2} f(\epsilon).
\end{eqnarray}
To get the total electronic population $N$, we calculate (for the BCME) 
\begin{eqnarray} \label{eq:eleNpes}
N = \int dx dp  \: \tilde{P}_1(x,p,t);
\end{eqnarray}
For the electronic friction-Langevin dynamics (EF-LD, Eq. \ref{eq:EF-LD}), we average the local population $n(h(x))$, 
\begin{eqnarray} \label{eq:LDN}
N= \langle n(h(x)) \rangle = \int dx dp \:  P_{LD} (x, p, t) n(h(x))
\end{eqnarray}
where $P_{LD} (x, p, t)$ is the total probability densities in phase space at position $x$ and $p$ from EF-LD.


Now, as far as friction is concerned, following  Ref. \citen{paperV}, one can show that Eqs. \ref{eq:cme3}-\ref{eq:cme4} are consistent with a
electronic friction of the form
\begin{eqnarray} \label{eq:ungammac}
\gamma_c (x) =  \frac{\hbar}{\Gamma}  \frac1{kT}    f(h(x)) \big(1-f(h(x))\big) (\partial_x h)^2, 
\end{eqnarray}  
Eq. \ref{eq:ungammac} is an {\em unbroadened} version of the friction term  $\gamma_1(x)$ (in Eq. \ref{eq:gamma1}). Including the effect of broadening
on friction is discussed in detail in  Ref. \citen{paperV}, where we have shown that such broadening effects do not
usually affect the dynamics very much; the effect of broadening on the potential of mean force surface is far stronger.

Finally, we must emphasize that Eqs. \ref{eq:cme3}-\ref{eq:cme4} do not incorporate any non-Condon effects with regards to frictional damping.
Thus, the terms $\gamma_2(x),\gamma_3(x), \gamma_4(x)$ in Eq. \ref{eq:gamma} are completely absent from our bCME in Eqs. \ref{eq:cme3}-\ref{eq:cme4}. While we would like to include these additional frictional terms, it is difficult to do so in a stable and easy manner because there is no guarantee that $\gamma_2(x)+\gamma_3(x)+\gamma_4(x)$ is greater than zero. All we are guaranteed is that $\gamma_1(x)+\gamma_2(x)+\gamma_3(x)+\gamma_4(x)>0$. See Eq. \ref{eq:fritotal}.  


\section{results} \label{sec:results}

Let us now apply the theory above to a simple model problem which extends the Anderson-Holstein model beyond the Condon approximation. 
For this problem, looking at Eq. \ref{Hs} and Eq. \ref{eq:Gamma}, we set 
\begin{eqnarray}
 U(x) &= & \frac12 m\omega^2 x^2, \label{eq:AHU}\\
 h(x) &=& E_d+ g \sqrt{{2m\omega}/{\hbar}} x, \label{eq:AHh} \\
\Gamma(x) &=& \Gamma_0 (1 +\exp(- K m \omega x^2/{\hbar} )) . \label{eq:AHgamma}  
\end{eqnarray}

\subsection{Statics}
\begin{figure}[htbp]
   \centering
   \scalebox{0.4}{\includegraphics{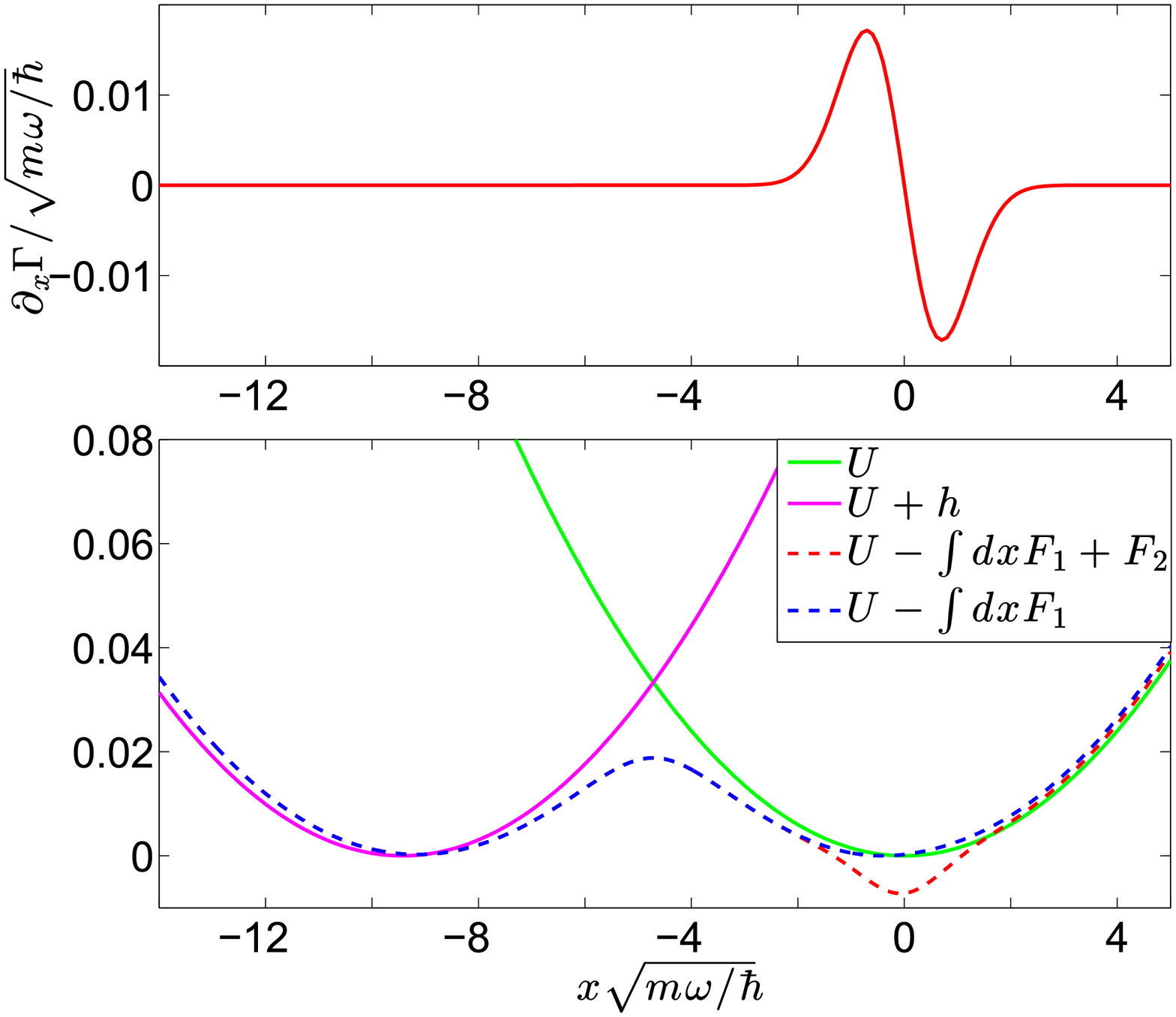} } 
   \caption{$\partial_x \Gamma$ (up) and Diabatic potentials as well as potentials of mean force (down) for the non-Condon Hamiltonian in Eqs. \ref{eq:AHU}-\ref{eq:AHgamma}:  $K=1$, $\hbar\omega=0.003$, $\Gamma_0=0.02$, $kT=0.01$, $g=0.02$, $\bar{E}_d=0$, $\mu=0$, $W=1$. $\bar{E}_d=E_d-g^2/\hbar\omega$ is the renormalized energy level. Notice that the presence of $F_2$ coming from $\partial_x \Gamma$ strongly modifies the potential of mean force.}
   \label{fig:pes}
\end{figure}

In Fig. \ref{fig:pes}, we plot the potentials of mean force (as well as the diabatic potentials $U(x)$ and $U(x)+h(x)$) as a function of nuclear position,
and we consider explicitly the effect of $F_2(x)$ (compare Eq. \ref{eq:Upmf1} with Eq. \ref{eq:Upmf2}).  From  Eq. \ref{eq:F2}, we know that $F_2(x)$ will distort the potential of mean force in regions where $\partial_x \Gamma$ is large.  This distortion
of the potential of mean force can dramatically affect the dynamical and equilibrium electronic population. Interestingly, in Fig. \ref{fig:pes}, we find that the potential of  mean force shows a dip in the region where $\Gamma$ has a peak, which indicates that the nuclei are attracted to positions of space where they hop back and forth frequently. This effect can be quantified by integrating Eq. \ref{eq:F2}. Suppose, for example, the integral $\int_{-W}^W \frac{d\epsilon}{2\pi}   (\epsilon-h(x)) A(\epsilon, x)  f(\epsilon)=-\alpha$ does not strongly depends on x (the integral is negative, so that $\alpha>0$). Then, the potential of mean force coming from the $F_2(x)$ term (see Eq. \ref{eq:Upmf2}) will be $U_{pmf}(x)=-\alpha \log(\Gamma(x))$, which indeed creates a dip where $\Gamma(x)$ is peaked. 

\begin{figure}[htbp]
   \centering
   \scalebox{0.5}{\includegraphics{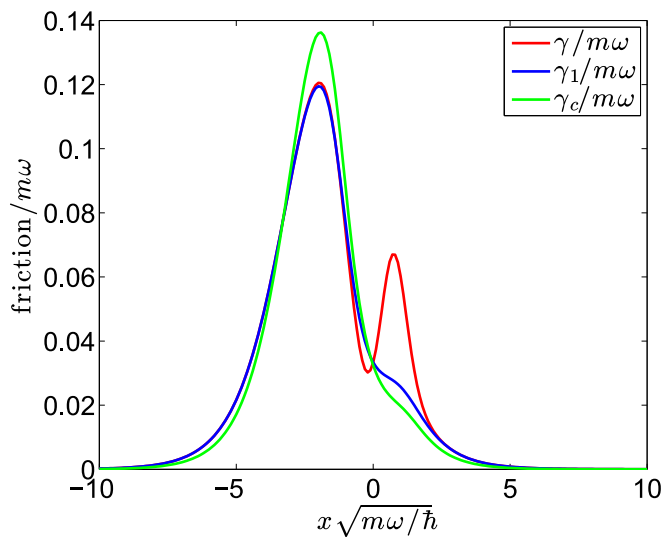} } 
   \caption{Three different approximations for electronic friction. $K=1$, $\hbar\omega=0.003$, $\Gamma_0=0.02$, $kT=0.01$, $g=0.0075$, $\bar{E}_d=0$, $\mu=0$. The total friction $\gamma(x) = \gamma_1(x)+\gamma_2(x)+\gamma_3(x)+\gamma_4(x)$ (Eq. \ref{eq:fritotal}) appears bimodal because of a  dip around $x=0$ where  $\partial_x \Gamma$ is large. This dip is not present either for  $\gamma_1(x)$ (Eq. \ref{eq:gamma1}) or $\gamma_c(x)$ (Eq. \ref{eq:ungammac}, i.e. the friction incorporated in the bCME).  }
   \label{fig:friction}
\end{figure}

In Fig. \ref{fig:friction}, we plot the electronic friction as a function of nuclear position. We do this for three cases: $\gamma(x)=\gamma_1(x)+\gamma_2(x)+\gamma_3(x)+\gamma_4(x)$, $\gamma_1(x)$ and $\gamma_c(x)$. Here $\gamma_c(x)$ is the CME friction (Eq. \ref{eq:ungammac}), which is the unbroadened version of $\gamma_1(x)$ (Eq. \ref{eq:gamma1}), and neither $\gamma_1(x)$ or $\gamma_c(x)$ includes any terms dependent on $\partial_x \Gamma$ ($\gamma_2(x), \gamma_3(x), \gamma_4(x)$).  Note that $\gamma_1(x)$ and $\gamma_c(x)$ have only one maximum  where the two PES's cross and the nuclei hop back and forth most frequently. The total friction $\gamma(x)$ is bimodal
because of a dip around $x=0$, where $\partial_x \Gamma$ is large (see Eq.  \ref{eq:fritotal}). For all three cases,  $D(x)=2kT\gamma(x)$ holds.  Thus, we may expect that all three cases give the same equilibrium electronic population and nuclear distribution (as long as the friction is not zero).

\subsection{Dynamics}
We now compare both electronic and nuclear dynamics (electronic population and kinetic energy as a function of time) from $(i)$ our bCME (Eqs. \ref{eq:cme3}-\ref{eq:cme4}) and $(ii)$ electronic friction-Langevin dynamics (EF-LD, Eq. \ref{eq:EF-LD}).
For EF-LD, the nuclei
simply move along the adiabatic potential of mean force (Eq. \ref{eq:decompose}) and feel friction $\gamma(x)$ (Eq. \ref{eq:gamma}, Eq. \ref{eq:fritotal}) and a random force $\delta f(x,t)$ (Eq. \ref{eq:corrDtotalso}, Eq. \ref{eq:corrtotal}). Thus, we emphasize that EF-LD dynamics correctly incorporate all non-Condon frictional components. 

For both algorithms, we initialize dynamics
with the nuclei equilibrated as a Gaussian distribution with a initial temperature $5kT$ and centered at position $x_1=-\sqrt{2\hbar/m\omega}g/\hbar\omega$. 
For the bCME, we initialize the electronic state for the molecule as being occupied, $N=1$. For EF-LD, the electronic population is always evaluated by averaging the local population $n(h(x))$  (Eq. \ref{eq:nh}) using the positon $x$ of each  trajectory (Eq. \ref{eq:LDN}).  

\begin{figure}[htbp]
   \centering
   \scalebox{0.5}{\includegraphics{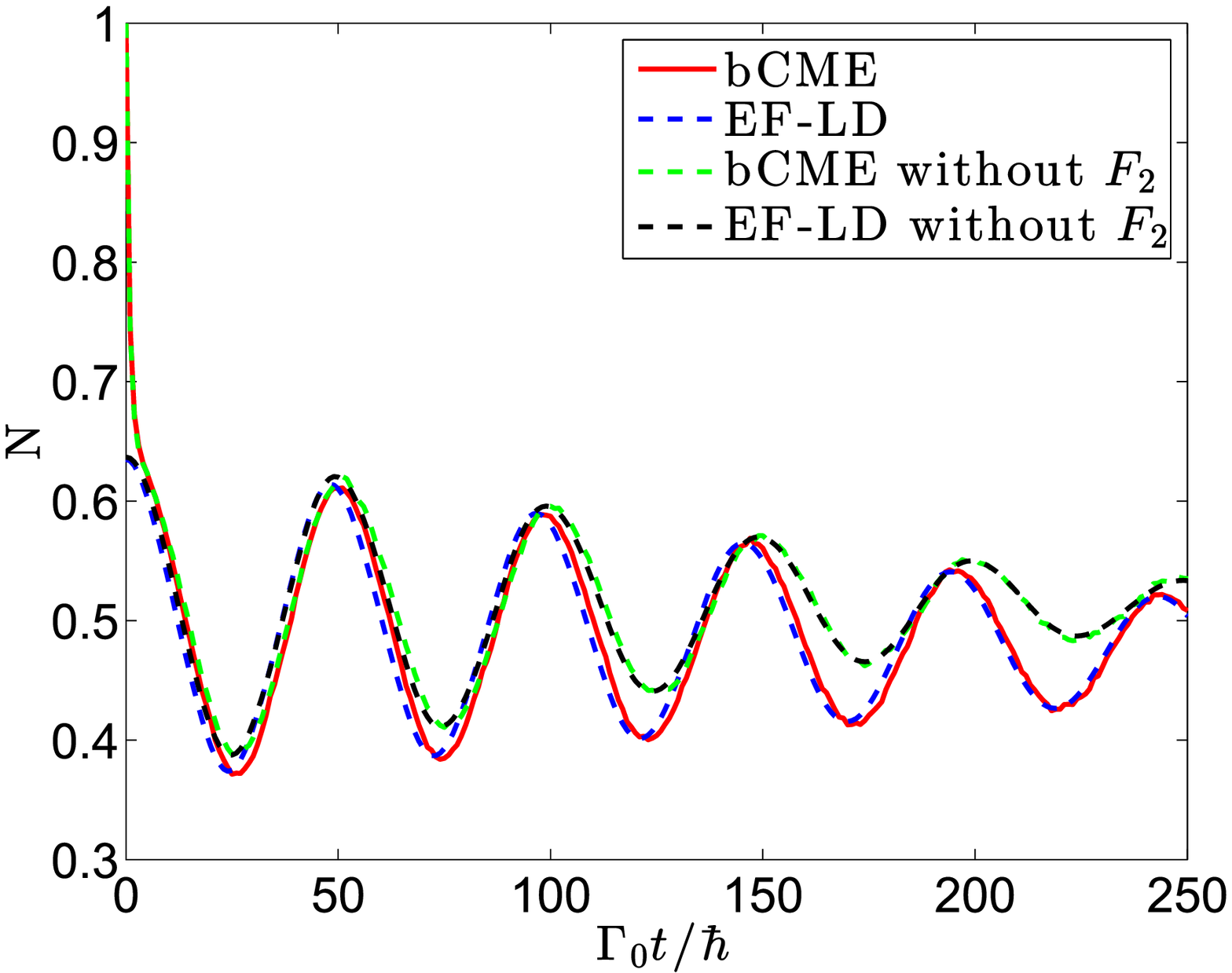} } 
   \caption{Electronic population N as a function of time for the generalized AH model in Eqs. \ref{eq:ah1}-\ref{eq:ah4}: $K=1$, $\hbar\omega=0.003$, $\Gamma_0=0.02$, $kT=0.01$, $g=0.0075$, $\bar{E}_d=0$, $\mu=0$, $W=1$. See Eqs. \ref{eq:AHU}-\ref{eq:AHgamma} for definition of the parameters.  And see Eqs. \ref{eq:eleNpes}-\ref{eq:LDN} for the definition of the electronic population $N$. Note that our bCME starts from correct initial conditions and agrees with EF-LD at later time. The mean force $F_2(x)$ affects the electronic population dramatically at long times. bCME (Eqs. \ref{eq:cme3}-\ref{eq:cme4}), EF-LD (Eq. \ref{eq:EF-LD} with $F(x)=F_1(x)+F_2(x)$),  bCME without $F_2(x)$ (Eqs. \ref{eq:cme1}-\ref{eq:cme2}), EF-LD without $F_2(x)$ (Eq. \ref{eq:EF-LD}, with $F(x)=F_1(x)$). In both cases, EF-LD dynamics include all of the contributions to the total non-Condon friction $\gamma(x)=\gamma_1(x)+\gamma_2(x)+\gamma_3(x)+\gamma_4(x)$. }
   \label{fig:ele}
\end{figure}

As Fig. \ref{fig:ele} shows, our bCME can recover the correct initial electronic population ($N=1$, see Ref. \citen{paperV}), whereas EF-LD cannot. As expected, at longer time, the bCME does agree with EF-LD. In the absence of any non-Condon contributions to
the potential mean force (i.e. $F_2(x)$), both the bCME and EF-LD reach an incorrect steady electronic population. Hence, it is essential to include the extra mean force ($F_2(x)$) arising from $\partial_x \Gamma$ into any dynamics. The results here are consistent with our observations regarding Fig. \ref{fig:pes}, where the contribution of $F_2(x)$ yields a significant dip in the region around x=0. 

\begin{figure}[htbp]
    \centering
     \subfloat[]{{\includegraphics[width=9cm]{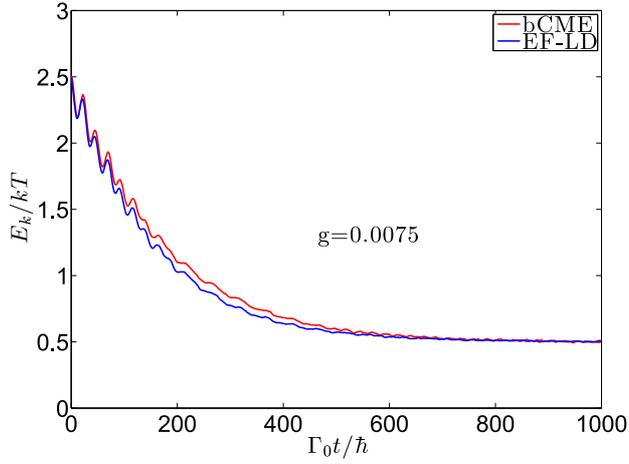} }}%
     \\
     \subfloat[]{{\includegraphics[width=9cm]{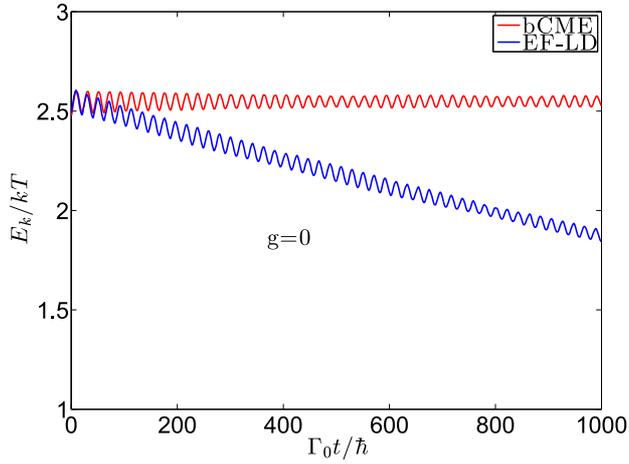} }}
    \caption{Average kinetic energy as a function of time for the generalized AH model in Eqs. \ref{eq:ah1}-\ref{eq:ah4}: $K=1$, $\hbar\omega=0.003$, $\Gamma_0=0.02$, $kT=0.01$, $\bar{E}_d=0$, $\mu=0$, $W=1$. See Eqs. \ref{eq:AHU}-\ref{eq:AHgamma} for definition of the parameters. (a) For $g=0.0075$, we get good agreement between bCME and EF-LD, even though the bCME friction (Eq. \ref{eq:fritotal}) is different from total EF-LD friction (Eq. \ref{eq:ungammac}). (b) However, bCME does fail when $\partial_x h$ is very small. Here, $\partial_x h=\sqrt{2}g=0$ so that the kinetic energy does not relax at all according to the bCME.  }
   \label{fig:Ek} 
\end{figure}

Finally, in Fig. \ref{fig:Ek}, we plot the average kinetic energy of the nuclei as a function of time for both the bCME and EF-LD. The relaxation rate for the nuclear motion is a measure of the amount friction. When $\partial_x h$ is not too small
(Fig. \ref{fig:Ek}(a), $g=0.0075$), we find good agreement between the bCME and EF-LD dynamics, even though our bCME friction is different from EF-LD total friction (see Fig. \ref{fig:friction}).  Generally speaking, we see an overall larger friction in EF-LD (see Fig. \ref{fig:friction}), which results in a slightly faster relaxation rate in Fig. \ref{fig:Ek}(a). By contrast, if we take the extreme case that $g=0$ so that  $\partial_x h=0$,
the frictional damping terms for  bCME and EF-LD are extremely different. In such a case,  as Fig. \ref{fig:Ek} shows, we see very large differences in the nuclear dynamics between bCME and EF-LD.  

In practice,  we anticipate that $\partial_x h$ will rarely be zero globally and so we cannot be 
sure how important such frictional effects will be.  In fact, for a condensed phase problem, it is possible that other sources
of friction from the environment may well overwhelm all of the effects of electronic friction.  These questions
will be addressed in future applications studies. \cite{FTforAddFriction}


\section{conclusion} \label{sec:con}
We have derived explicit forms for the electronic friction and random force from the generalized Anderson-Holstein (AH)
model in the case that the Condon approximation is violated ($\partial_x \Gamma \ne 0$) -- provided that the electronic motion is much faster than the nuclear motion (i.e. large $\Gamma$).  At equilibrium, the friction and random force satisfy the fluctuation-dissipation theorem. Our results can be generalized to the case of many nuclear degrees of freedom (see Appendix \ref{app:manyDoF}).
These results should be very useful in simulating frictional dynamics near metal surfaces in the adiabatic limit.
In general, our simulations show that violating the Condon approximation can dramatically affect both the dynamics and the equilibrium distribution.  

Focusing on dynamics, we  have shown how to incorporate the extra mean force coming from $\partial_x \Gamma$ into a broadened classical master equation (bCME).  After incorporating that extra mean force, our bCME agrees much better with electronic-friction langevin dynamics (EF-LD) in the adiabatic regime.  However, our proposed bCME does not incorporate the effect of $\partial_x \Gamma$
on the random force and friction, and thus will fail when $\partial_x h$ is much smaller than $\partial_x \Gamma$.  Further work will
explore approaches to incorporate these additional frictional forces.

\section{acknowledgments}
 We thank Abe Nitzan for very useful conversations.
This material is based upon work supported by the (U.S.) Air Force
Office of Scientific Research (USAFOSR) PECASE award under AFOSR Grant
No. FA9950-13-1-0157. J.E.S. acknowledges a Cottrell Research Scholar
Fellowship and a David and Lucille Packard Fellowship.

\section{Appendix: Details of the calculations }
 In the Appendix, we provide additional details of the calculations for friction and random force, we generalize our results to the case of many nuclear DoFs, we compare the result from two bCMEs,   and we establish a connection between our model and the Head-Gordon/Tully (HGT) model. For shorthand, we do not include dependence on $\epsilon$ or $x$ for functions. Thus, we write $f\equiv f(\epsilon)$, $A\equiv A(\epsilon, x)$, etc.

\subsection{Evaluating Friction} \label{app:a}
In this Appendix, we evaluate all frictional terms explicitly.  We first look at the $\gamma_1$ term (Eq. \ref{eq:gamma1raw}). Knowing the frozen Green's function exactly, one can derive Eq. \ref{eq:gamma1} by repeatedly integrating by parts, 
\begin{eqnarray} 
\gamma_1 &=& \hbar (\partial_x h)^2 \int \frac{d\epsilon}{2\pi} [ \partial_{\epsilon} G^< (G^A-G^R)]  \nonumber \\
&=&\hbar (\partial_x h)^2 \int \frac{d\epsilon}{2\pi} [ \partial_{\epsilon} (iAf) (iA) ] =\hbar (\partial_x h)^2 \int \frac{d\epsilon}{2\pi} Af \partial_{\epsilon} A \nonumber \\
 &=&\hbar (\partial_x h)^2 \frac12 \int \frac{d\epsilon}{2\pi} f \partial_{\epsilon} A^2 = - (\partial_x h)^2 \frac{\hbar}2 \int \frac{d\epsilon}{2\pi} A^2 \partial_{\epsilon} f. 
\end{eqnarray}

For $\gamma_2$, from Eq. \ref{eq:gamma2raw}, we have 
\begin{eqnarray} \label{eq:gamma2again}
\gamma_2 &=& (\partial_x h \partial_x \Gamma)   \frac{i\hbar}2 \int \frac{d\epsilon}{2\pi} [\partial_{\epsilon} G^< (G^A+G^R)+ (\partial_{\epsilon} G^R G^A-  G^R \partial_{\epsilon} G^A) f ] \nonumber \\
&=& - \hbar (\partial_x h \partial_x \Gamma)   \int \frac{d\epsilon}{2\pi} [ \partial_{\epsilon} (Af) \frac{A(\epsilon-h)}{\Gamma} + \frac{A^2}{2\Gamma} f ] . 
\end{eqnarray}
Again, we use integration by parts repeatedly for the first term on the right hand side of the above equation, 
\begin{eqnarray}  \label{eq:gamma2first}
 &&\int \frac{d\epsilon}{2\pi} \partial_{\epsilon} (Af) A(\epsilon-h) 
 = - \int \frac{d\epsilon}{2\pi}  Af  \partial_{\epsilon} [ A(\epsilon-h) ] \nonumber \\
 &=&  - \int \frac{d\epsilon}{2\pi} \Big( A^2f +  Af  (\epsilon-h) \partial_{\epsilon} A  \Big) 
=  - \int \frac{d\epsilon}{2\pi} \Big( A^2f + \frac12  f  (\epsilon-h) \partial_{\epsilon} A^2   \Big) \nonumber   \\
 &=&  - \int \frac{d\epsilon}{2\pi} \Big( A^2f - \frac12 A^2  \partial_{\epsilon} [f  (\epsilon-h) ]  \Big) 
 = - \frac12 \int \frac{d\epsilon}{2\pi} \Big( A^2f - A^2 (\epsilon-h)  \partial_{\epsilon} f    \Big) . 
\end{eqnarray}
Plugging Eq. \ref{eq:gamma2first} back into Eq. \ref{eq:gamma2again}, we arrive at a compact form of $\gamma_2$ (Eq. \ref{eq:gamma2})

To construct $\gamma_3$, we must recall that $\partial_x f = 0$ (of course). Then, if we evaluate the terms, 
\begin{eqnarray}
&&\sum_k   {V}_k^2 \Re ( \partial_x G^R \partial_{\epsilon} g_k^< +  \partial_x G^< \partial_{\epsilon} g_k^a ) \nonumber \\
= && \sum_k {V}_k^2  \Big(   \partial_x (-\frac i 2 A) \partial_{\epsilon} \big(i2\pi \delta(\epsilon-\epsilon_k) f \big) +   \partial_x (iAf) \partial_{\epsilon} (i\pi \delta(\epsilon-\epsilon_k) ) \Big) \nonumber \\
=&&  \pi \sum_k {V}_k^2  \delta(\epsilon-\epsilon_k) \partial_x A \partial_{\epsilon} f = \frac{\Gamma}2 \partial_x A \partial_{\epsilon} f . 
\end{eqnarray}
$\gamma_3$ (Eq. \ref{eq:gamma3}) eventually becomes 
\begin{eqnarray}
\gamma_3
&=& -\frac{\hbar \partial_x \Gamma} {4}  \int \frac{d\epsilon}{2\pi} \partial_x A \partial_{\epsilon} f \nonumber \\
&=& -  \frac{\hbar (\partial_x \Gamma)^2} 4 \int \frac{d\epsilon}{2\pi} (\frac{A}{\Gamma}- \frac{A^2}2) \partial_{\epsilon} f 
-  \frac{\hbar(\partial_x \Gamma)(\partial_x h) } 2 \int \frac{d\epsilon}{2\pi} \frac{A^2}{\Gamma} (\epsilon-h) \partial_{\epsilon} f. 
\end{eqnarray}

Similarly, $\gamma_4$ can be expressed as  
\begin{eqnarray}
\gamma_4 &=& -\frac{\hbar\partial_x \Gamma} {2\Gamma} \sum_k   {V}_k  \partial_x   {V}_k  \int \frac{d\epsilon}{2\pi} 2\Re (   G^R \partial_{\epsilon} g_k^< +  G^< \partial_{\epsilon} g_k^a) \nonumber \\
&=&  -\frac{\hbar \partial_x \Gamma} {2\Gamma}  \sum_k  \partial_ x    {V}_k^2  \int \frac{d\epsilon}{2\pi} \pi A  \delta(\epsilon-\epsilon_k)  \partial_{\epsilon} f  \nonumber \\
&=&  -\frac{\hbar \partial_x \Gamma} {2\Gamma}   \frac{\partial_ x \Gamma}2  \int \frac{d\epsilon}{2\pi}  A  \partial_{\epsilon} f . 
\end{eqnarray}

\subsection{Evaluating the Correlation Functions for the Random Force} \label{app:b}
Evaluating the correlation functions for the random force is very similar to evaluating the current noise for a resonant model and can be found, for example, in Ref. \citen{negf}. To evaluate the correlation function, we work in the energy domain, 
\begin{eqnarray} \label{eq:D12energy}
D_{12}  &=&  \frac { \partial_x h \partial_x \Gamma}  {2\Gamma} 2\hbar \Re \sum_k  {V}_k \int  \frac{d\epsilon}{2\pi} G_{d,k}^>  G^<  , \\
\label{eq:D21energy}
D_{21} &=&  \frac { \partial_x h \partial_x \Gamma}  {2\Gamma} 2\hbar \Re \sum_k  {V}_k \int  \frac{d\epsilon}{2\pi} G^>  G_{d,k}^<  ,  \\
\label{eq:D22energy}
D_{22} &=& \left( \frac{\partial_x \Gamma } {2\Gamma} \right ) ^2 2\hbar \Re\sum_{k,k'}   {V}_k   {V}_{k'} \int \frac{d\epsilon}{2\pi}   G_{k,d}^> G^<_{k',d} \nonumber \\
&+& \left( \frac{\partial_x \Gamma } {2\Gamma} \right ) ^2 2\hbar \Re\sum_{k,k'}   {V}_k   {V}_{k'} \int \frac{d\epsilon}{2\pi}   G_{k,k'}^> G^<. 
\end{eqnarray}
We then evaluate the following terms by using the Langreth decomposition, 
\begin{eqnarray} \label{eq:corrGlessgreat}
\sum_k  {V}_k  G^{< }_{d,k} &=& \sum_k V_k^2 ( G^R g_k^{< } + G^{< }  g_k^a ) \nonumber \\
&=& G^R \Sigma^{<} + G^{< } \Sigma^A = if A(\epsilon-h) , 
\end{eqnarray} 
Similarly, one can show that 
\begin{eqnarray} \label{eq:corrGklessgreat}
\sum_k  {V}_k  G^{> }_{d,k} &=&  -i(1-f) A(\epsilon-h),  \\
\sum_k  {V}_k  G^{< }_{k,d} &=& i f A(\epsilon-h),  \\
\sum_k  {V}_k  G^{> }_{k,d} &=& -i (1-f) A(\epsilon-h). 
\end{eqnarray} 
We also need to evaluate terms such as
\begin{eqnarray}  \label{eq:corrVVG}
&& \sum_{k,k'}   {V}_k   {V}_{k'} G^>_{k,k'}  \nonumber \\
&=& \sum_{k,k'}   {V}_k    {V}_{k'} \delta_{k,k'} g_k^>
+   {V}_k^2   {V}_{k'}^2  [g_{k}^r G^R g_{k'}^> + g_{k}^r G^> g_{k'}^a+g_{k}^> G^A g_{k'}^a] \nonumber \\
&=& \Sigma^> + \Sigma^R G^R \Sigma^> + \Sigma^R G^> \Sigma^A + \Sigma^> G^A \Sigma^A  \nonumber \\
&=&-i(1-f)\Big( \Gamma + (-i \Gamma/2) G^R \Gamma  + A (\Gamma/2)^2 + \Gamma G^A (i \Gamma/2) \Big) \nonumber \\
&=&-i(1-f) \Big( \Gamma - \frac{\Gamma^2}4 A \Big) = -i(1-f)  A (\epsilon-h)^2 .  
\end{eqnarray}
Plugging Eqs. \ref{eq:corrGlessgreat}-\ref{eq:corrVVG} into Eqs. \ref{eq:D12energy}-\ref{eq:D22energy}, one can easily get Eq. \ref{eq:D1221s}-\ref{eq:D22total}. 

\subsection{Multiple nuclear degrees of freedom} \label{app:manyDoF}
For $N$ nuclear degrees of freedom, the system Hamiltonian and the interaction Hamiltonian from Eq. \ref{Hs} and Eq. \ref{eq:ah4} become:
\begin{eqnarray}
 {H}_s &=& h(x_1,...,x_N)  {d}^+ {d} + \sum_{\alpha=1}^{N}\frac{p_{\alpha}^2}{2m_{\alpha} } + U(x_1,...,x_N), \\
 {H}_c &=& \sum_k {V_k}(x_1,...,x_N) (  {d}^+  {c}_k +  {c}_k^+  {d} ). 
\end{eqnarray}
One can follow the exact derivation as in the main body of this paper and show that the resulting Langevin equation becomes
\begin{eqnarray} 
-m_{\alpha} \ddot x_{\alpha}  =\partial_{x_{\alpha} }U  - F_{\alpha} + \sum_{\beta} \gamma_{\alpha \beta} \dot{x}_{\beta} + \delta f_{\alpha} (t), 
\end{eqnarray}
where the mean force is
\begin{eqnarray}
F_{\alpha} =  -  \int_{-W}^W \frac{d\epsilon}{2\pi} \left(\partial_{x_{\alpha} } h + (\epsilon-h)\frac{\partial_{x_{\alpha} } \Gamma}{\Gamma} \right)  A f .  
\end{eqnarray}
and friction is 
\begin{eqnarray}
\gamma_{\alpha \beta } =  - \frac{\hbar}{2} \int \frac{d\epsilon}{2\pi} \left( \partial_{x_{\alpha} } h + (\epsilon-h)\frac{\partial_{x_{\alpha} } \Gamma}{\Gamma} \right)   \left( \partial_{x_{\beta} } h + (\epsilon-h)\frac{\partial_{x_{\beta} } \Gamma}{\Gamma} \right)  A^2\partial_{\epsilon}  f. 
\end{eqnarray}
The random force again is Markovian,  $\langle \delta f_{\alpha} (t) f_{\beta} (t') \rangle = D_{\alpha \beta } \delta(t-t')$, with 
\begin{eqnarray} 
D_{\alpha \beta }= \hbar \int \frac{d\epsilon}{2\pi} \left( \partial_{x_{\alpha} } h + (\epsilon-h)\frac{\partial_{x_{\alpha} } \Gamma}{\Gamma} \right) \left( \partial_{x_{\beta} } h + (\epsilon-h)\frac{\partial_{x_{\beta} } \Gamma}{\Gamma} \right) A^2  f(1- f) . 
\end{eqnarray}

\subsection{A comparison of two bCMEs} \label{app:c}
\begin{figure}[htbp]
   \centering
   \scalebox{0.4}{\includegraphics{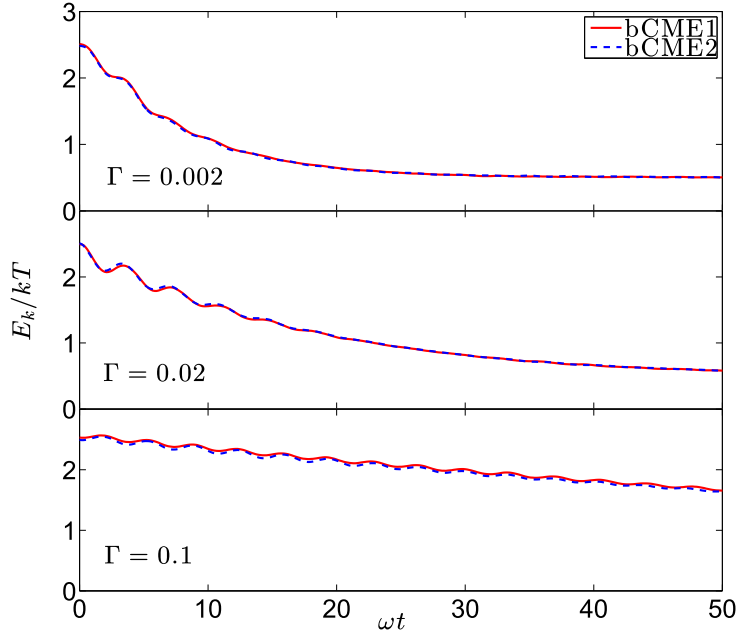} } 
   \caption{Average kinetic energy as a function of time: $\hbar\omega=0.003$, $kT=0.01$, $g=0.01$, $\bar{E}_d=0$, $\mu=0$. bCME1 refers to Eqs. \ref{eq:bCMEori1}-\ref{eq:bCMEori2}. bCME2 refers to Eqs. \ref{eq:cme1}-\ref{eq:cme2}. Note the near perfect agreement between the two bCMEs. }
   \label{fig:bCMEtwo}
\end{figure}

In Ref. \citen{paperV}, we previously used a slightly different bCME to incorporate level broadening. The bCME in Ref. \citen{paperV} (which we refer to as bCME1) reads: 
\begin{eqnarray} \label{eq:bCMEori1}
\frac { \partial P_0(x,p,t)} {\partial t} &=&  - \frac{p}{m}  \frac {\partial P_0(x,p,t)} {\partial x} + \partial_x U  \frac {\partial P_0(x,p,t)} { \partial p} \nonumber \\
&-&\frac{\Gamma}{\hbar} f(h) P_0(x,p,t)  
+ \frac{\Gamma}{\hbar} \big(1-f(h)\big) P_1(x,p,t) \nonumber \\
&+& \big(- F_1(x) -f(h) \partial_x h \big) (1-f(h))  \frac {\partial \big(P_0(x,p,t) + P_1(x,p,t) \big)} { \partial p} ,
\end{eqnarray}
\begin{eqnarray}  \label{eq:bCMEori2}
\frac { \partial P_1(x,p,t)} {\partial t} &=&  - \frac{p}{m}  \frac {\partial P_1 (x,p,t)} {\partial x} +  \Big(\partial_x U +  \partial_x h  \Big) \frac {\partial P_1 (x,p,t)} { \partial p}  \nonumber \\
&+& \frac{\Gamma}{\hbar} f(h) P_0(x,p,t) 
- \frac{\Gamma}{\hbar} \big(1-f(h)\big) P_1(x,p,t) \nonumber \\
&+& \big(- F_1(x) -f(h) \partial_x h\big)  f(h)  \frac {\partial \big( P_0(x,p,t) + P_1(x,p,t) \big) } { \partial p} . 
\end{eqnarray}
Eqs. \ref{eq:bCMEori1}-\ref{eq:bCMEori2} work well for a constant $\Gamma$ (i.e. the Condon approximation). Comparing this bCME with the alternate bCME we are using in the main body of the paper (bCME2, Eqs. \ref{eq:cme1}-\ref{eq:cme2}), we notice that momentum jumps are required to solve bCME1 (Eqs. \ref{eq:bCMEori1}-\ref{eq:bCMEori2}) with trajectories (because $\partial P_0/\partial t$ ($\partial P_1/\partial t$) includes $\partial P_1/\partial p$ ($\partial P_0/\partial p$) ). However, momentum jumps are not present in bCME2 (Eqs. \ref{eq:cme1}-\ref{eq:cme2}).
Obviously, because we  have constructed our bCMEs by extrapolation from the diabatic limit to the adiabatic limit, we cannot expect to find along a single unique set of equations. That being said, because the momentum jump is only a first order approximation for  solving  a series of entangled partial differential equations, we may expect momentum jump solutions may fail for very large $g$. By contract, bCME2 should be still trustworthy even for very large $g$. Thus, we have worked with bCME2 in the present paper.  Moveover, Fig. \ref{fig:bCMEtwo} shows these two bCMEs agree with each other for a large range of parameters.

\subsection{Head-Gordon and Tully friction model}  \label{app:HGT}
Previously, in Ref. \citen{paperIV}, we argued that there is a disconnect between our  frictional model and the HGT model when we go from a finite system to a manifold of 
electronic states. At this point, however, we will show that a natural connection can be constructed if one extrapolates the HGT model properly to the limit of infinitely many electronic states. 
For the HGT model, the electronic friction is given by \cite{FrictionTully, shenvi:2009:iesh}, 
\begin{eqnarray} \label{eq:HGTfriction}
\gamma =\pi\hbar  d^2_{l, l'} = \pi\hbar 
\frac{| \langle l | {\partial_x H_e }| l' \rangle |^2 } {(\epsilon_{l}-\epsilon_{l'})^2  }, 
\end{eqnarray}
where $| l \rangle$ ($| l' \rangle$) is the adiabatic state just below (above) the Fermi level. We have used Hellmann-Feynman theorem in the last equality with  the electronic Hamiltonian $H_e$ defined as 
\begin{eqnarray}
H_e= h(x) d^+ d + \sum_k \epsilon_k c^+_k c_k + \sum_k V_k(x) ( d^+ c_k + c^+_k d ). 
\end{eqnarray}
In the context of infinite electronic DoFs, the HGT friction is  \cite{FrictionTully}
\begin{eqnarray} \label{eq:tullyFriction0T}
\gamma  = \pi \hbar (\langle l | \partial_x H_e| l \rangle)^2 \rho^2(\epsilon_l)|_{\epsilon_l = \epsilon_F}. 
\end{eqnarray}  
Here $\rho(\epsilon_l)$ is the density of states $| l \rangle $ with an energy $\epsilon_l$. $\epsilon_F$ is the Fermi level.

We note that the HGT model was derived for zero temperature. We propose that,  at finite temperature, the natural extension of the HGT model should be  
\begin{eqnarray} \label{eq:tullyFrictionFT}
\gamma = -\pi \hbar \int d\epsilon_l \partial_{\epsilon_l} f(\epsilon_l) (\langle l | \partial_x H_e| l \rangle)^2 \rho^2(\epsilon_l). 
\end{eqnarray}  
At zero temperature, Eq. \ref{eq:tullyFrictionFT} reduces to Eq. \ref{eq:tullyFriction0T} by noting $-\partial_{\epsilon_l} f(\epsilon_l)=\delta(\epsilon_l-\epsilon_F)$. Now we will show that Eq. \ref{eq:tullyFrictionFT} is exactly the same as what we derived in the main body of the paper. 

We first evaluate the term, 
\begin{eqnarray}
\langle l | \partial_x  H_e  | l \rangle = \langle l | \partial_x h \: d^+ d + \sum_k \partial_x {V_k} (  {d}^+  {c}_k +  {c}_k^+  {d} )| l \rangle.  
\end{eqnarray}                                                                                                                                                                                                                                                                                                                                                                                                                                                                                                                                                                                                                                                                                                                                                                                                                                                                                                                                                                                                                                                                  
We proceed by expressing $H_e$  in a basis of adiabatic states, \cite{mahan, bijayprb2015} 
\begin{eqnarray}
H_e= \sum_l \epsilon_l C^+_l C_l, 
\end{eqnarray}
where
\begin{eqnarray}
d &=& \sum_l \alpha_l C_l,  \: \: \alpha_l = \frac{V_l} { \epsilon_l - h - \sum_{l'} \frac{V_{l'}^2}{ \epsilon_l -\epsilon_{l'}+ i \eta }}, \\
c_k &= &\sum_l \beta_{k l} C_l , \:\:  \beta_{kl} = \delta_{kl} - \frac{V_k \alpha_{l} } { \epsilon_k - \epsilon_{l} + i \eta} . 
\end{eqnarray}
Here, we apply the wide band approximation (Eq. \ref{eq:gamma}), such that $\alpha_l =  \frac{V_l} { \epsilon_l - h + i \Gamma/2} $. 

Now we are ready to evaluate  
\begin{eqnarray}
\langle l | d^+ d | l \rangle = \alpha_l^* \alpha_l = \frac{V_l^2}{(\epsilon_l -h)^2 + (\Gamma/2)^2} , 
\end{eqnarray}
and 
\begin{eqnarray}
&&\sum_k \partial_x {V_k}  \langle l |  (  {d}^+  {c}_k +  {c}_k^+  {d} )   | l \rangle  = 2 \sum_k \partial_x {V_k}\Re (\langle l |   {d}^+  {c}_k   | l \rangle ) 
= 2 \sum_k \partial_x {V_k}\Re (\alpha_l ^* \beta_{kl} ) \nonumber \\
&=& 2 \partial_x V_l \Re (\alpha_l^*) - 2 \alpha_l^* \alpha_l  \sum_k \Re \frac{V_k \partial_x V_k}  { \epsilon_k - \epsilon_{l} + i \eta} 
=  \frac{(\epsilon_l -h) \partial_x V_l^2} {(\epsilon_l - h)^2 + (\Gamma/2)^2} . 
\end{eqnarray}
In the last equality, we have assumed $\Re \sum_k \frac{\partial_x V_k^2 } { \epsilon_k - \epsilon_{l} + i \eta} =0 $ (as a result of the wide band approximation). Using $2\pi V_l^2 \rho(\epsilon_l)=\Gamma$ (and $2\pi \rho(\epsilon_l) \partial_x V_l^2 =\partial_x \Gamma$), and switching integration  variables from $\epsilon_l$ to $\epsilon$, we arrive at the same friction as Eq. \ref{eq:fritotal}, 
\begin{eqnarray} \label{eq:frictionagain}
\gamma 
=   \frac{\hbar}2 \int  \frac{d\epsilon}{2\pi} f(\epsilon)  (1-f(\epsilon)) 
 \frac{1}{kT} \left( \partial_x h  +  \frac{\partial_x \Gamma (\epsilon -h) } {\Gamma}    \right)^2 A^2(\epsilon, x). 
\end{eqnarray}

 Here we have used the fact that $\partial_{\epsilon} f(\epsilon)= -f(\epsilon)(1-f(\epsilon))/kT$. Thus, Eq. \ref{eq:tullyFrictionFT} is a suitable extension of HGT model to finite temperature that agrees with our (and von Oppen \textit{et al}'s \cite{beilstein}) picture of friction.  Note that Eq. \ref{eq:frictionagain} was derived previously in Refs. \citen{Mizielinski2005, Mizielinski2007, mishaPRBfriction}.


\begin{thebibliography}{42}%
\makeatletter
\providecommand \@ifxundefined [1]{%
 \@ifx{#1\undefined}
}%
\providecommand \@ifnum [1]{%
 \ifnum #1\expandafter \@firstoftwo
 \else \expandafter \@secondoftwo
 \fi
}%
\providecommand \@ifx [1]{%
 \ifx #1\expandafter \@firstoftwo
 \else \expandafter \@secondoftwo
 \fi
}%
\providecommand \natexlab [1]{#1}%
\providecommand \enquote  [1]{``#1''}%
\providecommand \bibnamefont  [1]{#1}%
\providecommand \bibfnamefont [1]{#1}%
\providecommand \citenamefont [1]{#1}%
\providecommand \href@noop [0]{\@secondoftwo}%
\providecommand \href [0]{\begingroup \@sanitize@url \@href}%
\providecommand \@href[1]{\@@startlink{#1}\@@href}%
\providecommand \@@href[1]{\endgroup#1\@@endlink}%
\providecommand \@sanitize@url [0]{\catcode `\\12\catcode `\$12\catcode
  `\&12\catcode `\#12\catcode `\^12\catcode `\_12\catcode `\%12\relax}%
\providecommand \@@startlink[1]{}%
\providecommand \@@endlink[0]{}%
\providecommand \url  [0]{\begingroup\@sanitize@url \@url }%
\providecommand \@url [1]{\endgroup\@href {#1}{\urlprefix }}%
\providecommand \urlprefix  [0]{URL }%
\providecommand \Eprint [0]{\href }%
\providecommand \doibase [0]{http://dx.doi.org/}%
\providecommand \selectlanguage [0]{\@gobble}%
\providecommand \bibinfo  [0]{\@secondoftwo}%
\providecommand \bibfield  [0]{\@secondoftwo}%
\providecommand \translation [1]{[#1]}%
\providecommand \BibitemOpen [0]{}%
\providecommand \bibitemStop [0]{}%
\providecommand \bibitemNoStop [0]{.\EOS\space}%
\providecommand \EOS [0]{\spacefactor3000\relax}%
\providecommand \BibitemShut  [1]{\csname bibitem#1\endcsname}%
\let\auto@bib@innerbib\@empty
\bibitem [{\citenamefont {Huang}\ \emph {et~al.}(2000)\citenamefont {Huang},
  \citenamefont {Rettner}, \citenamefont {Auerbach},\ and\ \citenamefont
  {Wodtke}}]{science2000}%
  \BibitemOpen
  \bibfield  {author} {\bibinfo {author} {\bibfnamefont {Y.}~\bibnamefont
  {Huang}}, \bibinfo {author} {\bibfnamefont {C.~T.}\ \bibnamefont {Rettner}},
  \bibinfo {author} {\bibfnamefont {D.~J.}\ \bibnamefont {Auerbach}}, \ and\
  \bibinfo {author} {\bibfnamefont {A.~M.}\ \bibnamefont {Wodtke}},\
  }\href@noop {} {\bibfield  {journal} {\bibinfo  {journal} {{\em Science}}\
  }\textbf {\bibinfo {volume} {290}},\ \bibinfo {pages} {111} (\bibinfo {year}
  {2000})}\BibitemShut {NoStop}%
\bibitem [{\citenamefont {Bartels}\ \emph {et~al.}(2011)\citenamefont
  {Bartels}, \citenamefont {Cooper}, \citenamefont {Auerbach},\ and\
  \citenamefont {Wodtke}}]{wodtke}%
  \BibitemOpen
  \bibfield  {author} {\bibinfo {author} {\bibfnamefont {C.}~\bibnamefont
  {Bartels}}, \bibinfo {author} {\bibfnamefont {R.}~\bibnamefont {Cooper}},
  \bibinfo {author} {\bibfnamefont {D.~J.}\ \bibnamefont {Auerbach}}, \ and\
  \bibinfo {author} {\bibfnamefont {A.~M.}\ \bibnamefont {Wodtke}},\
  }\href@noop {} {\bibfield  {journal} {\bibinfo  {journal} {{\em Chem. Sci.}}\
  }\textbf {\bibinfo {volume} {2}},\ \bibinfo {pages} {1647} (\bibinfo {year}
  {2011})}\BibitemShut {NoStop}%
\bibitem [{\citenamefont {Shenvi}, \citenamefont {Roy},\ and\ \citenamefont
  {Tully}(2009{\natexlab{a}})}]{shenvi:2009:iesh}%
  \BibitemOpen
  \bibfield  {author} {\bibinfo {author} {\bibfnamefont {N.}~\bibnamefont
  {Shenvi}}, \bibinfo {author} {\bibfnamefont {S.}~\bibnamefont {Roy}}, \ and\
  \bibinfo {author} {\bibfnamefont {J.~C.}\ \bibnamefont {Tully}},\ }\href@noop
  {} {\bibfield  {journal} {\bibinfo  {journal} {{\em J. Chem. Phys.}}\
  }\textbf {\bibinfo {volume} {130}},\ \bibinfo {pages} {174107} (\bibinfo
  {year} {2009}{\natexlab{a}})}\BibitemShut {NoStop}%
\bibitem [{\citenamefont {Shenvi}, \citenamefont {Roy},\ and\ \citenamefont
  {Tully}(2009{\natexlab{b}})}]{shenvi:2009:science}%
  \BibitemOpen
  \bibfield  {author} {\bibinfo {author} {\bibfnamefont {N.}~\bibnamefont
  {Shenvi}}, \bibinfo {author} {\bibfnamefont {S.}~\bibnamefont {Roy}}, \ and\
  \bibinfo {author} {\bibfnamefont {J.~C.}\ \bibnamefont {Tully}},\ }\href@noop
  {} {\bibfield  {journal} {\bibinfo  {journal} {{\em Science}}\ }\textbf
  {\bibinfo {volume} {326}},\ \bibinfo {pages} {829} (\bibinfo {year}
  {2009}{\natexlab{b}})}\BibitemShut {NoStop}%
\bibitem [{\citenamefont {Galperin}, \citenamefont {Ratner},\ and\
  \citenamefont {Nitzan}(2007)}]{inelastic}%
  \BibitemOpen
  \bibfield  {author} {\bibinfo {author} {\bibfnamefont {M.}~\bibnamefont
  {Galperin}}, \bibinfo {author} {\bibfnamefont {M.~A.}\ \bibnamefont
  {Ratner}}, \ and\ \bibinfo {author} {\bibfnamefont {A.}~\bibnamefont
  {Nitzan}},\ }\href@noop {} {\bibfield  {journal} {\bibinfo  {journal} {{\em
  J. Phys: Condens. Matter}}\ }\textbf {\bibinfo {volume} {19}},\ \bibinfo
  {pages} {103201} (\bibinfo {year} {2007})}\BibitemShut {NoStop}%
\bibitem [{\citenamefont {M{\"u}hlbacher}\ and\ \citenamefont
  {Rabani}(2008)}]{prldata}%
  \BibitemOpen
  \bibfield  {author} {\bibinfo {author} {\bibfnamefont {L.}~\bibnamefont
  {M{\"u}hlbacher}}\ and\ \bibinfo {author} {\bibfnamefont {E.}~\bibnamefont
  {Rabani}},\ }\href@noop {} {\bibfield  {journal} {\bibinfo  {journal} {{\em
  Phys. Rev. Lett.}}\ }\textbf {\bibinfo {volume} {100}},\ \bibinfo {pages}
  {176403} (\bibinfo {year} {2008})}\BibitemShut {NoStop}%
\bibitem [{\citenamefont {McEniry}\ \emph {et~al.}(2007)\citenamefont
  {McEniry}, \citenamefont {Bowler}, \citenamefont {Dundas}, \citenamefont
  {Horsfield}, \citenamefont {S\'{a}nchez},\ and\ \citenamefont
  {Todorov}}]{bowler2007}%
  \BibitemOpen
  \bibfield  {author} {\bibinfo {author} {\bibfnamefont {E.~J.}\ \bibnamefont
  {McEniry}}, \bibinfo {author} {\bibfnamefont {D.~R.}\ \bibnamefont {Bowler}},
  \bibinfo {author} {\bibfnamefont {D.}~\bibnamefont {Dundas}}, \bibinfo
  {author} {\bibfnamefont {A.~P.}\ \bibnamefont {Horsfield}}, \bibinfo {author}
  {\bibfnamefont {C.~G.}\ \bibnamefont {S\'{a}nchez}}, \ and\ \bibinfo {author}
  {\bibfnamefont {T.~N.}\ \bibnamefont {Todorov}},\ }\href@noop {} {\bibfield
  {journal} {\bibinfo  {journal} {{\em J. Phys: Condens. Matter}}\ }\textbf
  {\bibinfo {volume} {19}},\ \bibinfo {pages} {196201} (\bibinfo {year}
  {2007})}\BibitemShut {NoStop}%
\bibitem [{\citenamefont {Galperin}\ \emph {et~al.}(2008)\citenamefont
  {Galperin}, \citenamefont {Ratner}, \citenamefont {Nitzan},\ and\
  \citenamefont {Troisi}}]{mishaScience2008}%
  \BibitemOpen
  \bibfield  {author} {\bibinfo {author} {\bibfnamefont {M.}~\bibnamefont
  {Galperin}}, \bibinfo {author} {\bibfnamefont {M.~A.}\ \bibnamefont
  {Ratner}}, \bibinfo {author} {\bibfnamefont {A.}~\bibnamefont {Nitzan}}, \
  and\ \bibinfo {author} {\bibfnamefont {A.}~\bibnamefont {Troisi}},\
  }\href@noop {} {\bibfield  {journal} {\bibinfo  {journal} {Science}\ }\textbf
  {\bibinfo {volume} {319}},\ \bibinfo {pages} {1056} (\bibinfo {year}
  {2008})}\BibitemShut {NoStop}%
\bibitem [{\citenamefont {Galperin}, \citenamefont {Ratner},\ and\
  \citenamefont {Nitzan}(2005)}]{Galperinnano}%
  \BibitemOpen
  \bibfield  {author} {\bibinfo {author} {\bibfnamefont {M.}~\bibnamefont
  {Galperin}}, \bibinfo {author} {\bibfnamefont {M.~A.}\ \bibnamefont
  {Ratner}}, \ and\ \bibinfo {author} {\bibfnamefont {A.}~\bibnamefont
  {Nitzan}},\ }\href@noop {} {\bibfield  {journal} {\bibinfo  {journal} {{\em
  Nano Lett.}}\ }\textbf {\bibinfo {volume} {5}},\ \bibinfo {pages} {125}
  (\bibinfo {year} {2005})}\BibitemShut {NoStop}%
\bibitem [{\citenamefont {Joachim}\ and\ \citenamefont
  {Ratner}(2005)}]{rater2005review}%
  \BibitemOpen
  \bibfield  {author} {\bibinfo {author} {\bibfnamefont {C.}~\bibnamefont
  {Joachim}}\ and\ \bibinfo {author} {\bibfnamefont {M.~A.}\ \bibnamefont
  {Ratner}},\ }\href@noop {} {\bibfield  {journal} {\bibinfo  {journal} {Proc.
  Nat. Acad. Sci. USA}\ }\textbf {\bibinfo {volume} {102}},\ \bibinfo {pages}
  {8801} (\bibinfo {year} {2005})}\BibitemShut {NoStop}%
\bibitem [{\citenamefont {Wu}\ \emph {et~al.}(2008)\citenamefont {Wu},
  \citenamefont {Ogawa}, \citenamefont {Nazin},\ and\ \citenamefont
  {Ho}}]{hysertesisJPCC}%
  \BibitemOpen
  \bibfield  {author} {\bibinfo {author} {\bibfnamefont {S.~W.}\ \bibnamefont
  {Wu}}, \bibinfo {author} {\bibfnamefont {N.}~\bibnamefont {Ogawa}}, \bibinfo
  {author} {\bibfnamefont {G.~V.}\ \bibnamefont {Nazin}}, \ and\ \bibinfo
  {author} {\bibfnamefont {W.}~\bibnamefont {Ho}},\ }\href@noop {} {\bibfield
  {journal} {\bibinfo  {journal} {{\em J. Phys. Chem. C}}\ }\textbf {\bibinfo
  {volume} {112}},\ \bibinfo {pages} {5241} (\bibinfo {year}
  {2008})}\BibitemShut {NoStop}%
\bibitem [{\citenamefont {L\"{o}rtscher}\ \emph {et~al.}(2006)\citenamefont
  {L\"{o}rtscher}, \citenamefont {Ciszek}, \citenamefont {Tour},\ and\
  \citenamefont {Riel}}]{switching2006}%
  \BibitemOpen
  \bibfield  {author} {\bibinfo {author} {\bibfnamefont {E.}~\bibnamefont
  {L\"{o}rtscher}}, \bibinfo {author} {\bibfnamefont {J.}~\bibnamefont
  {Ciszek}}, \bibinfo {author} {\bibfnamefont {J.}~\bibnamefont {Tour}}, \ and\
  \bibinfo {author} {\bibfnamefont {H.}~\bibnamefont {Riel}},\ }\href@noop {}
  {\bibfield  {journal} {\bibinfo  {journal} {Small}\ }\textbf {\bibinfo
  {volume} {2}},\ \bibinfo {pages} {973} (\bibinfo {year} {2006})}\BibitemShut
  {NoStop}%
\bibitem [{\citenamefont {Koch}\ \emph {et~al.}(2006)\citenamefont {Koch},
  \citenamefont {Semmelhack}, \citenamefont {von Oppen},\ and\ \citenamefont
  {Nitzan}}]{prbnitzan}%
  \BibitemOpen
  \bibfield  {author} {\bibinfo {author} {\bibfnamefont {J.}~\bibnamefont
  {Koch}}, \bibinfo {author} {\bibfnamefont {M.}~\bibnamefont {Semmelhack}},
  \bibinfo {author} {\bibfnamefont {F.}~\bibnamefont {von Oppen}}, \ and\
  \bibinfo {author} {\bibfnamefont {A.}~\bibnamefont {Nitzan}},\ }\href@noop {}
  {\bibfield  {journal} {\bibinfo  {journal} {{\em Phys. Rev. B}}\ }\textbf
  {\bibinfo {volume} {73}},\ \bibinfo {pages} {155306} (\bibinfo {year}
  {2006})}\BibitemShut {NoStop}%
\bibitem [{\citenamefont {Kaasbjerg}, \citenamefont {Novotn{\' y}},\ and\
  \citenamefont {Nitzan}(2013)}]{heating}%
  \BibitemOpen
  \bibfield  {author} {\bibinfo {author} {\bibfnamefont {K.}~\bibnamefont
  {Kaasbjerg}}, \bibinfo {author} {\bibfnamefont {T.}~\bibnamefont {Novotn{\'
  y}}}, \ and\ \bibinfo {author} {\bibfnamefont {A.}~\bibnamefont {Nitzan}},\
  }\href@noop {} {\bibfield  {journal} {\bibinfo  {journal} {{\em Phys. Rev.
  B}}\ }\textbf {\bibinfo {volume} {88}},\ \bibinfo {pages} {201405} (\bibinfo
  {year} {2013})}\BibitemShut {NoStop}%
\bibitem [{\citenamefont {H\"{a}rtle}\ and\ \citenamefont
  {Thoss}(2011)}]{thossinstabilities}%
  \BibitemOpen
  \bibfield  {author} {\bibinfo {author} {\bibfnamefont {R.}~\bibnamefont
  {H\"{a}rtle}}\ and\ \bibinfo {author} {\bibfnamefont {M.}~\bibnamefont
  {Thoss}},\ }\href@noop {} {\bibfield  {journal} {\bibinfo  {journal} {{\em
  Phys. Rev. B}}\ }\textbf {\bibinfo {volume} {83}},\ \bibinfo {pages} {125419}
  (\bibinfo {year} {2011})}\BibitemShut {NoStop}%
\bibitem [{\citenamefont {Arrachea}, \citenamefont {Bode},\ and\ \citenamefont
  {von Oppen}(2014)}]{cooling}%
  \BibitemOpen
  \bibfield  {author} {\bibinfo {author} {\bibfnamefont {L.}~\bibnamefont
  {Arrachea}}, \bibinfo {author} {\bibfnamefont {N.}~\bibnamefont {Bode}}, \
  and\ \bibinfo {author} {\bibfnamefont {F.}~\bibnamefont {von Oppen}},\
  }\href@noop {} {\bibfield  {journal} {\bibinfo  {journal} {{\em Phys. Rev.
  B}}\ }\textbf {\bibinfo {volume} {90}},\ \bibinfo {pages} {125450} (\bibinfo
  {year} {2014})}\BibitemShut {NoStop}%
\bibitem [{\citenamefont {Tully}(1980)}]{tully1980}%
  \BibitemOpen
  \bibfield  {author} {\bibinfo {author} {\bibfnamefont {J.~C.}\ \bibnamefont
  {Tully}},\ }\href@noop {} {\bibfield  {journal} {\bibinfo  {journal} {{\em J.
  Chem. Phys.}}\ }\textbf {\bibinfo {volume} {73}},\ \bibinfo {pages} {1975}
  (\bibinfo {year} {1980})}\BibitemShut {NoStop}%
\bibitem [{\citenamefont {Adelman}\ and\ \citenamefont
  {Doll}(1976)}]{adelman1976}%
  \BibitemOpen
  \bibfield  {author} {\bibinfo {author} {\bibfnamefont {S.~A.}\ \bibnamefont
  {Adelman}}\ and\ \bibinfo {author} {\bibfnamefont {J.~D.}\ \bibnamefont
  {Doll}},\ }\href@noop {} {\bibfield  {journal} {\bibinfo  {journal} {{\em J.
  Chem. Phys.}}\ }\textbf {\bibinfo {volume} {64}},\ \bibinfo {pages} {2375}
  (\bibinfo {year} {1976})}\BibitemShut {NoStop}%
\bibitem [{\citenamefont {Head-Gordon}\ and\ \citenamefont
  {Tully}(1995)}]{FrictionTully}%
  \BibitemOpen
  \bibfield  {author} {\bibinfo {author} {\bibfnamefont {M.}~\bibnamefont
  {Head-Gordon}}\ and\ \bibinfo {author} {\bibfnamefont {J.~C.}\ \bibnamefont
  {Tully}},\ }\href@noop {} {\bibfield  {journal} {\bibinfo  {journal} {{\em J.
  Chem. Phys.}}\ }\textbf {\bibinfo {volume} {103}},\ \bibinfo {pages} {10137}
  (\bibinfo {year} {1995})}\BibitemShut {NoStop}%
\bibitem [{\citenamefont {Struck}\ \emph {et~al.}(1996)\citenamefont {Struck},
  \citenamefont {Richter}, \citenamefont {Buntin}, \citenamefont {Cavanagh},\
  and\ \citenamefont {Stephenson}}]{tullyfrictionexpPRL1996}%
  \BibitemOpen
  \bibfield  {author} {\bibinfo {author} {\bibfnamefont {L.~M.}\ \bibnamefont
  {Struck}}, \bibinfo {author} {\bibfnamefont {L.~J.}\ \bibnamefont {Richter}},
  \bibinfo {author} {\bibfnamefont {S.~A.}\ \bibnamefont {Buntin}}, \bibinfo
  {author} {\bibfnamefont {R.~R.}\ \bibnamefont {Cavanagh}}, \ and\ \bibinfo
  {author} {\bibfnamefont {J.~C.}\ \bibnamefont {Stephenson}},\ }\href@noop {}
  {\bibfield  {journal} {\bibinfo  {journal} {{\em Phys. Rev. Lett.}}\ }\textbf
  {\bibinfo {volume} {77}},\ \bibinfo {pages} {4576} (\bibinfo {year}
  {1996})}\BibitemShut {NoStop}%
\bibitem [{\citenamefont {F\"{u}chsel}\ \emph {et~al.}(2011)\citenamefont
  {F\"{u}chsel}, \citenamefont {Klamroth}, \citenamefont {Montureta},\ and\
  \citenamefont {Saalfrank}}]{tullyappPCCP2011}%
  \BibitemOpen
  \bibfield  {author} {\bibinfo {author} {\bibfnamefont {G.}~\bibnamefont
  {F\"{u}chsel}}, \bibinfo {author} {\bibfnamefont {T.}~\bibnamefont
  {Klamroth}}, \bibinfo {author} {\bibfnamefont {S.}~\bibnamefont {Montureta}},
  \ and\ \bibinfo {author} {\bibfnamefont {P.}~\bibnamefont {Saalfrank}},\
  }\href@noop {} {\bibfield  {journal} {\bibinfo  {journal} {{\em Phys. Chem.
  Chem. Phys.}}\ }\textbf {\bibinfo {volume} {13}},\ \bibinfo {pages} {8659}
  (\bibinfo {year} {2011})}\BibitemShut {NoStop}%
\bibitem [{\citenamefont {Tully}(2000)}]{tullyfrictionreview}%
  \BibitemOpen
  \bibfield  {author} {\bibinfo {author} {\bibfnamefont {J.~C.}\ \bibnamefont
  {Tully}},\ }\href@noop {} {\bibfield  {journal} {\bibinfo  {journal} {Annual
  Review of Physical Chemistry}\ }\textbf {\bibinfo {volume} {51}},\ \bibinfo
  {pages} {153} (\bibinfo {year} {2000})}\BibitemShut {NoStop}%
\bibitem [{\citenamefont {Wodtke}, \citenamefont {Tully},\ and\ \citenamefont
  {Auerbach}(2004)}]{wodtkereview2004}%
  \BibitemOpen
  \bibfield  {author} {\bibinfo {author} {\bibfnamefont {A.~M.}\ \bibnamefont
  {Wodtke}}, \bibinfo {author} {\bibfnamefont {J.~C.}\ \bibnamefont {Tully}}, \
  and\ \bibinfo {author} {\bibfnamefont {D.~J.}\ \bibnamefont {Auerbach}},\
  }\href@noop {} {\bibfield  {journal} {\bibinfo  {journal} {International
  Reviews in Physical Chemistry}\ }\textbf {\bibinfo {volume} {23}},\ \bibinfo
  {pages} {513} (\bibinfo {year} {2004})}\BibitemShut {NoStop}%
\bibitem [{\citenamefont {Bode}\ \emph {et~al.}(2012)\citenamefont {Bode},
  \citenamefont {Kusminskiy}, \citenamefont {Egger},\ and\ \citenamefont {von
  Oppen}}]{beilstein}%
  \BibitemOpen
  \bibfield  {author} {\bibinfo {author} {\bibfnamefont {N.}~\bibnamefont
  {Bode}}, \bibinfo {author} {\bibfnamefont {S.~V.}\ \bibnamefont
  {Kusminskiy}}, \bibinfo {author} {\bibfnamefont {R.}~\bibnamefont {Egger}}, \
  and\ \bibinfo {author} {\bibfnamefont {F.}~\bibnamefont {von Oppen}},\
  }\href@noop {} {\bibfield  {journal} {\bibinfo  {journal} {{\em Beilstein J.
  Nanotechnol}}\ }\textbf {\bibinfo {volume} {3}},\ \bibinfo {pages} {144}
  (\bibinfo {year} {2012})}\BibitemShut {NoStop}%
\bibitem [{\citenamefont {Thomas}\ \emph {et~al.}(2012)\citenamefont {Thomas},
  \citenamefont {Karzig}, \citenamefont {Kusminskiy}, \citenamefont
  {Zar\'{a}nd},\ and\ \citenamefont {von Oppen}}]{vonOppenPRB}%
  \BibitemOpen
  \bibfield  {author} {\bibinfo {author} {\bibfnamefont {M.}~\bibnamefont
  {Thomas}}, \bibinfo {author} {\bibfnamefont {T.}~\bibnamefont {Karzig}},
  \bibinfo {author} {\bibfnamefont {S.~V.}\ \bibnamefont {Kusminskiy}},
  \bibinfo {author} {\bibfnamefont {G.}~\bibnamefont {Zar\'{a}nd}}, \ and\
  \bibinfo {author} {\bibfnamefont {F.}~\bibnamefont {von Oppen}},\ }\href@noop
  {} {\bibfield  {journal} {\bibinfo  {journal} {{\em Phys. Rev. B}}\ }\textbf
  {\bibinfo {volume} {86}},\ \bibinfo {pages} {195419} (\bibinfo {year}
  {2012})}\BibitemShut {NoStop}%
\bibitem [{\citenamefont {Brandbyge}\ \emph {et~al.}(1995)\citenamefont
  {Brandbyge}, \citenamefont {Hedeg{\aa}rd}, \citenamefont {Heinz},
  \citenamefont {Misewich},\ and\ \citenamefont {Newns}}]{brandbyge}%
  \BibitemOpen
  \bibfield  {author} {\bibinfo {author} {\bibfnamefont {M.}~\bibnamefont
  {Brandbyge}}, \bibinfo {author} {\bibfnamefont {P.}~\bibnamefont
  {Hedeg{\aa}rd}}, \bibinfo {author} {\bibfnamefont {T.~F.}\ \bibnamefont
  {Heinz}}, \bibinfo {author} {\bibfnamefont {J.~A.}\ \bibnamefont {Misewich}},
  \ and\ \bibinfo {author} {\bibfnamefont {D.~M.}\ \bibnamefont {Newns}},\
  }\href@noop {} {\bibfield  {journal} {\bibinfo  {journal} {{\em Phys. Rev.
  B}}\ }\textbf {\bibinfo {volume} {52}},\ \bibinfo {pages} {6042} (\bibinfo
  {year} {1995})}\BibitemShut {NoStop}%
\bibitem [{\citenamefont {Mozyrsky}, \citenamefont {Hastings},\ and\
  \citenamefont {Martin}(2006)}]{Mozyrsky}%
  \BibitemOpen
  \bibfield  {author} {\bibinfo {author} {\bibfnamefont {D.}~\bibnamefont
  {Mozyrsky}}, \bibinfo {author} {\bibfnamefont {M.~B.}\ \bibnamefont
  {Hastings}}, \ and\ \bibinfo {author} {\bibfnamefont {I.}~\bibnamefont
  {Martin}},\ }\href@noop {} {\bibfield  {journal} {\bibinfo  {journal} {{\em
  Phys. Rev. B}}\ }\textbf {\bibinfo {volume} {73}},\ \bibinfo {pages} {035104}
  (\bibinfo {year} {2006})}\BibitemShut {NoStop}%
\bibitem [{\citenamefont {L\"{u}}\ \emph {et~al.}(2012)\citenamefont {L\"{u}},
  \citenamefont {Brandbyge}, \citenamefont {Hedeg{\aa}rd}, \citenamefont
  {Todorov},\ and\ \citenamefont {Dundas}}]{lvPRBfriction}%
  \BibitemOpen
  \bibfield  {author} {\bibinfo {author} {\bibfnamefont {J.-T.}\ \bibnamefont
  {L\"{u}}}, \bibinfo {author} {\bibfnamefont {M.}~\bibnamefont {Brandbyge}},
  \bibinfo {author} {\bibfnamefont {P.}~\bibnamefont {Hedeg{\aa}rd}}, \bibinfo
  {author} {\bibfnamefont {T.~N.}\ \bibnamefont {Todorov}}, \ and\ \bibinfo
  {author} {\bibfnamefont {D.}~\bibnamefont {Dundas}},\ }\href@noop {}
  {\bibfield  {journal} {\bibinfo  {journal} {{\em Phys. Rev. B}}\ }\textbf
  {\bibinfo {volume} {85}},\ \bibinfo {pages} {245444} (\bibinfo {year}
  {2012})}\BibitemShut {NoStop}%
\bibitem [{\citenamefont {Dou}, \citenamefont {Nitzan},\ and\ \citenamefont
  {Subotnik}(2015{\natexlab{a}})}]{paperIV}%
  \BibitemOpen
  \bibfield  {author} {\bibinfo {author} {\bibfnamefont {W.}~\bibnamefont
  {Dou}}, \bibinfo {author} {\bibfnamefont {A.}~\bibnamefont {Nitzan}}, \ and\
  \bibinfo {author} {\bibfnamefont {J.~E.}\ \bibnamefont {Subotnik}},\
  }\href@noop {} {\bibfield  {journal} {\bibinfo  {journal} {{\em J. Chem.
  Phys.}}\ }\textbf {\bibinfo {volume} {143}},\ \bibinfo {pages} {054103}
  (\bibinfo {year} {2015}{\natexlab{a}})}\BibitemShut {NoStop}%
\bibitem [{\citenamefont {Mizielinski}\ \emph {et~al.}(2005)\citenamefont
  {Mizielinski}, \citenamefont {Bird}, \citenamefont {Persson},\ and\
  \citenamefont {Holloway}}]{Mizielinski2005}%
  \BibitemOpen
  \bibfield  {author} {\bibinfo {author} {\bibfnamefont {M.~S.}\ \bibnamefont
  {Mizielinski}}, \bibinfo {author} {\bibfnamefont {D.~M.}\ \bibnamefont
  {Bird}}, \bibinfo {author} {\bibfnamefont {M.}~\bibnamefont {Persson}}, \
  and\ \bibinfo {author} {\bibfnamefont {S.}~\bibnamefont {Holloway}},\
  }\href@noop {} {\bibfield  {journal} {\bibinfo  {journal} {{\em J. Chem.
  Phys.}}\ }\textbf {\bibinfo {volume} {122}},\ \bibinfo {pages} {084710}
  (\bibinfo {year} {2005})}\BibitemShut {NoStop}%
\bibitem [{\citenamefont {Mizielinski}\ \emph {et~al.}(2007)\citenamefont
  {Mizielinski}, \citenamefont {Bird}, \citenamefont {Persson},\ and\
  \citenamefont {Holloway}}]{Mizielinski2007}%
  \BibitemOpen
  \bibfield  {author} {\bibinfo {author} {\bibfnamefont {M.~S.}\ \bibnamefont
  {Mizielinski}}, \bibinfo {author} {\bibfnamefont {D.~M.}\ \bibnamefont
  {Bird}}, \bibinfo {author} {\bibfnamefont {M.}~\bibnamefont {Persson}}, \
  and\ \bibinfo {author} {\bibfnamefont {S.}~\bibnamefont {Holloway}},\
  }\href@noop {} {\bibfield  {journal} {\bibinfo  {journal} {{\em J. Chem.
  Phys.}}\ }\textbf {\bibinfo {volume} {126}},\ \bibinfo {pages} {034705}
  (\bibinfo {year} {2007})}\BibitemShut {NoStop}%
\bibitem [{\citenamefont {Esposito}, \citenamefont {Ochoa},\ and\ \citenamefont
  {Galperin}(2015)}]{mishaPRBfriction}%
  \BibitemOpen
  \bibfield  {author} {\bibinfo {author} {\bibfnamefont {M.}~\bibnamefont
  {Esposito}}, \bibinfo {author} {\bibfnamefont {M.~A.}\ \bibnamefont {Ochoa}},
  \ and\ \bibinfo {author} {\bibfnamefont {M.}~\bibnamefont {Galperin}},\
  }\href@noop {} {\bibfield  {journal} {\bibinfo  {journal} {{\em Phys. Rev.
  B}}\ }\textbf {\bibinfo {volume} {92}},\ \bibinfo {pages} {235440} (\bibinfo
  {year} {2015})}\BibitemShut {NoStop}%
\bibitem [{\citenamefont {Dou}, \citenamefont {Nitzan},\ and\ \citenamefont
  {Subotnik}(2015{\natexlab{b}})}]{paperII}%
  \BibitemOpen
  \bibfield  {author} {\bibinfo {author} {\bibfnamefont {W.}~\bibnamefont
  {Dou}}, \bibinfo {author} {\bibfnamefont {A.}~\bibnamefont {Nitzan}}, \ and\
  \bibinfo {author} {\bibfnamefont {J.~E.}\ \bibnamefont {Subotnik}},\
  }\href@noop {} {\bibfield  {journal} {\bibinfo  {journal} {{\em J. Chem.
  Phys.}}\ }\textbf {\bibinfo {volume} {142}},\ \bibinfo {pages} {084110}
  (\bibinfo {year} {2015}{\natexlab{b}})}\BibitemShut {NoStop}%
\bibitem [{\citenamefont {Dou}, \citenamefont {Nitzan},\ and\ \citenamefont
  {Subotnik}(2015{\natexlab{c}})}]{paperIII}%
  \BibitemOpen
  \bibfield  {author} {\bibinfo {author} {\bibfnamefont {W.}~\bibnamefont
  {Dou}}, \bibinfo {author} {\bibfnamefont {A.}~\bibnamefont {Nitzan}}, \ and\
  \bibinfo {author} {\bibfnamefont {J.~E.}\ \bibnamefont {Subotnik}},\
  }\href@noop {} {\bibfield  {journal} {\bibinfo  {journal} {{\em J. Chem.
  Phys.}}\ }\textbf {\bibinfo {volume} {142}},\ \bibinfo {pages} {234106}
  (\bibinfo {year} {2015}{\natexlab{c}})}\BibitemShut {NoStop}%
\bibitem [{\citenamefont {Mahan}(2000)}]{mahan}%
  \BibitemOpen
  \bibfield  {author} {\bibinfo {author} {\bibfnamefont {G.~D.}\ \bibnamefont
  {Mahan}},\ }\href@noop {} {\emph {\bibinfo {title} {Many-Particle Physics}}}\
  (\bibinfo  {publisher} {Plenum},\ \bibinfo {address} {New York},\ \bibinfo
  {year} {2000})\BibitemShut {NoStop}%
\bibitem [{\citenamefont {Jauho}(2016)}]{JH1877notes}%
  \BibitemOpen
  \bibfield  {author} {\bibinfo {author} {\bibfnamefont {A.~P.}\ \bibnamefont
  {Jauho}},\ }\href {https://nanohub.org/resources/1877} {\enquote {\bibinfo
  {title} {Introduction to the keldysh nonequilibrium green function
  technique},}\ } (\bibinfo {year} {2016})\BibitemShut {NoStop}%
\bibitem [{\citenamefont {Tannor}(2006)}]{Tannor}%
  \BibitemOpen
  \bibfield  {author} {\bibinfo {author} {\bibfnamefont {D.~J.}\ \bibnamefont
  {Tannor}},\ }\href@noop {} {\emph {\bibinfo {title} {Introduction to quantum
  mechanics: a time-dependent perspective}}}\ (\bibinfo  {publisher}
  {University Science Books},\ \bibinfo {year} {2006})\BibitemShut {NoStop}%
\bibitem [{\citenamefont {Dou}, \citenamefont {Nitzan},\ and\ \citenamefont
  {Subotnik}(2016)}]{paperVI}%
  \BibitemOpen
  \bibfield  {author} {\bibinfo {author} {\bibfnamefont {W.}~\bibnamefont
  {Dou}}, \bibinfo {author} {\bibfnamefont {A.}~\bibnamefont {Nitzan}}, \ and\
  \bibinfo {author} {\bibfnamefont {J.~E.}\ \bibnamefont {Subotnik}},\
  }\href@noop {} {\bibfield  {journal} {\bibinfo  {journal} {{\em J. Chem.
  Phys.}}\ }\textbf {\bibinfo {volume} {144}},\ \bibinfo {pages} {074109}
  (\bibinfo {year} {2016})}\BibitemShut {NoStop}%
\bibitem [{\citenamefont {Dou}\ and\ \citenamefont {Subotnik}(2016)}]{paperV}%
  \BibitemOpen
  \bibfield  {author} {\bibinfo {author} {\bibfnamefont {W.}~\bibnamefont
  {Dou}}\ and\ \bibinfo {author} {\bibfnamefont {J.~E.}\ \bibnamefont
  {Subotnik}},\ }\href@noop {} {\bibfield  {journal} {\bibinfo  {journal} {{\em
  J. Chem. Phys.}}\ }\textbf {\bibinfo {volume} {144}},\ \bibinfo {pages}
  {024116} (\bibinfo {year} {2016})}\BibitemShut {NoStop}%
\bibitem [{FTf()}]{FTforAddFriction}%
  \BibitemOpen
  \href@noop {} {}\bibinfo {note} {Note that, if we find that the effects of
  the non-Condon obeying frictional terms ($\gamma_2, \gamma_3, \gamma_4$) is
  large, we can always use the the approach in Ref. \citen{paperV} to add the
  complementary friction to update the bCME.}\BibitemShut {Stop}%
\bibitem [{\citenamefont {Haug}\ and\ \citenamefont {Jauho}(2007)}]{negf}%
  \BibitemOpen
  \bibfield  {author} {\bibinfo {author} {\bibfnamefont {H.}~\bibnamefont
  {Haug}}\ and\ \bibinfo {author} {\bibfnamefont {A.}~\bibnamefont {Jauho}},\
  }\href@noop {} {\emph {\bibinfo {title} {Quantum Kinetics in Transport and
  Optics of Semiconductors.}}}\ (\bibinfo  {publisher} {Springer},\ \bibinfo
  {address} {New York},\ \bibinfo {year} {2007})\BibitemShut {NoStop}%
\bibitem [{\citenamefont {Agarwalla}, \citenamefont {Jiang},\ and\
  \citenamefont {Segal}(2015)}]{bijayprb2015}%
  \BibitemOpen
  \bibfield  {author} {\bibinfo {author} {\bibfnamefont {B.~K.}\ \bibnamefont
  {Agarwalla}}, \bibinfo {author} {\bibfnamefont {J.-H.}\ \bibnamefont
  {Jiang}}, \ and\ \bibinfo {author} {\bibfnamefont {D.}~\bibnamefont
  {Segal}},\ }\href@noop {} {\bibfield  {journal} {\bibinfo  {journal} {{\em
  Phys. Rev. B}}\ }\textbf {\bibinfo {volume} {92}},\ \bibinfo {pages} {245418}
  (\bibinfo {year} {2015})}\BibitemShut {NoStop}%
\end{thebibliography}
\end{document}